\documentclass[a4paper,12pt]{article}
\usepackage{amssymb}

\usepackage{CJK}
\usepackage{amsmath}
\usepackage{graphicx}
\usepackage[caption=false,font=normalsize,labelfont=sf,textfont=sf]{subfig}
\usepackage{color}
\usepackage{url}

\newcommand{\eqnb}{\begin{equation}}
\newcommand{\eqne}{\end{equation}}

\newtheorem{The}{Theorem}

\newtheorem{Lem}{Lemma}

\newtheorem{Rem}{Remark}

\begin{document}

\title{A Markov Process Theory for Network Growth Processes of DAG-based
Blockchain Systems}
\author{Xing-Shuo Song$^{a}$, Quan-Lin Li$^{b}$, Yan-Xia Chang$^{b,}$\thanks{Corresponding author. E-mail address: changyanxia@emails.bjut.edu.cn. }, Chi Zhang$^{b}$ \\
$^{a}$School of Economics and Management\\
Yanshan University, Qinhuangdao 066004, China\\
$^{b}$School of Economics and Management\\
Beijing University of Technology, Beijing 100124, China}

\maketitle

\footnotetext{ This work has been submitted to the IEEE for possible publication. Copyright may be transferred without notice, after which this version may no longer be accessible.  }

\begin{abstract}
Note that the serial structure of blockchain has many essential
pitfalls, thus a data network structure and its DAG-based blockchain are
introduced to resolve the blockchain pitfalls. From such a network
perspective, analysis of the DAG-based blockchain systems becomes
interesting and challenging. So, the simulation models are
adopted widely. In this paper, we first describe a simple Markov model for the DAG-based
blockchain with IOTA Tangle by means of two layers of tips and internal
tips' impatient connection behavior. Then we set up a continuous-time Markov
process to analyze the DAG-based blockchain system and show that this Markov
process is a level-dependent quasi-birth-and-death (QBD) process. Based on
this, we prove that the QBD process must be irreducible and positive
recurrent. Furthermore, once the stationary probability vector of the QBD
process is given, we provide performance analysis of the DAG-based blockchain
system. Nextly, we propose a new effective method for computing the average
confirmation time of any arriving internal tip at this system by means of the first
passage times and the PH distributions. Finally, we use numerical examples
to check the validity of our theoretical results and indicate how some key
system parameters influence the performance measures of this system. Therefore, we hope that the
methodology and results developed in this paper can be applicable to deal with more general DAG-based
blockchain systems such that a series of promising research can be developed
potentially.

\end{abstract}

\section{Introduction}

Blockchain technologies originated from Bitcoin by Nakamoto \cite{Nak:2008} in
2008. Since then, blockchain has attracted tremendous attention from both
practitioners and academics in many different areas, such as finance,
reputation system, security, public service, anti-corruption, Internet of
Things, and so on. Blockchain has many remarkable and excellent features, for example, decentralization,
distributed structure, availability, persistency, consistency, anonymity,
immutability, auditability, and accountability. Bitcoin and Ethereum
are regarded as the two most representative blockchain technologies that
have primarily contributed to such popularity gain. So far, blockchain
technologies have significantly enabled a wide spectrum of applications
from ``cryptocurrency, financial service, reputation system,
security, public service, smart contracts, Internet of Things'' to
``healthcare, energy management, supply chain management, sharing economy,
social governance, insurance, law, art among others''. Readers may refer to
recent survey papers for details, among which are Wang et al. \cite
{Wan:2019b}, Gorkhali et al. \cite{Gor:2020}, Belchior et al. \cite{Bel:2021}
and Huang et al. \cite{Hua:2021}.

It is worthwhile to note that Bitcoin stores transactions in a publicly
available ledger. Each block contains a finite batch of transactions and
the blocks are strung together in a chronological order such that they can
construct a chain of blocks. The consistency of the ledger comes from the
structural extension in the series of blocks. The newly added
block is viewed as legitimate only if it is consistent with the chain of all
those blocks in front of it. After Bitcoin, other cryptocurrencies, like
Ethereum and Litecoin, originate from extending and generalizing the useful
functionalities of Bitcoin. But they still retain the block-to-block series
connection of physical structure. For Bitcoin, Ethereum and other blockchain
systems, when there is an honest mining pool (note that the multiple honest
mining pools can merge into a total honest mining pool) and multiple
different dishonest mining pools in the system, it can cause multiple block branches on a tree. In this case, the blockchain can be set up by pegging each main chain
at each round, which is a competitive result observed from a tree of blocks by means
of the longest chain principle. Notice that the backbone of this tree is
built by the honest mining pool, while the other branches are built by only
the dishonest mining pools under such a physical setting that a dishonest mining
pool can only build one branch at most. See Li et al. \cite{Li:2021, Li:2022}
for more details.

However, the serial structure of blockchain (including a blockchain
generated from the tree of blocks with multiple branches) has many essential pitfalls, such as poor performance and scalability, limited
transaction throughput, high transaction cost, long confirmation delay, huge
energy expenditure (Proof of Work), and so forth. To resolve these blockchain
pitfalls, a DAG data network structure was introduced to the blockchain
technologies (called a DAG-based blockchain) in order to replace the serial
structure of the blockchain. On the one hand, the DAG-based blockchain has
quickly emerged in the context of the Internet of Things, e.g., see Ferrag
et al. \cite{Fer:2018a}, Lo et al. \cite{Lo:2019}, Viriyasitavat et al. \cite%
{Vir:2019}, Cullen et al. \cite{Cul:2020} and Alshaikhli et al. \cite%
{Als:2021}. On the other hand, the DAG-based blockchain has increasingly
been adopted in the field of Distributed Ledger Technology, e.g., see Zhu et
al. \cite{Zhu:2019}, Gorbunova et al. \cite{Gor:2022}, Park et al. \cite%
{Par:2019} and Ben\v{c}i\'{c} and \v{Z}arko \cite{Ben:2018}.

To develop the DAG-based blockchain and facilitate secure payments
and communication between devices of the Internet of Things, the IOTA
Foundation proposed a new protocol called IOTA, which has a DAG data network
structure. Readers may refer to papers, such as Popov %
\cite{Pop:2016, Pop:2018}, Popov and Buchanan \cite{PopB:2019}, and Popov et
al. \cite{Pop:2020, Pop:2019}. From the Popov's papers, it is easy to see
that all the transactions would be approved and stored permanently on the
IOTA Tangle. In addition, to further understand the basic elements of the
IOTA Tangle, readers may refer to Conti et al. \cite{Con:2022}, Attias and
Bramas \cite{Att:2019}, and Fan \cite{Fan:2019} for more details.

So far, the Internet of Things has gained rapid development and has also been applied to many different practical fields,
such as logistics, supply chain management, healthcare, smart city,
intelligent transportation, intelligent security, intelligent building,
intelligent home, heritage protection, positioning and navigation, video
surveillance, and so on. It is well known that the Internet of Things will
generate massive amounts of data every day. In this case, blockchain is a
key technology for setting up a necessary and effective platform to store
and manage the data from the Internet of Things. Also, in this
research line, it is the key to apply the IOTA Tangle to the Internet of Things due
to the fact that the DAG-based blockchain has high transaction throughput,
good performance, and low cost. To this end, readers may refer to survey
papers by Alshaikhli et al. \cite{Als:2021} and Lo et al. \cite{Lo:2019};
Distributed Ledger Technology by Cullen et al. \cite{Cul:2019, Cul:2020},
Park et al. \cite{Par:2019} and Siim \cite{Sii:2018}; and industrial
Internet of Things by Cui et al. \cite{Cui:2019} and Liao et al. \cite%
{Lia:2022}.

Up to now, there has been some key research on the DAG-based blockchain with
IOTA Tangle. To help readers quickly understand the recent literature and
associated advances, here we provide\ an overview of DAG-based
blockchain as follows:

\textbf{(1)} The survey papers: Wang et al. \cite{Wan:2020}, Bai \cite%
{Bai:2018}, Siim \cite{Sii:2018}, Brunner \cite{Bru:2021} and Keidar et al. %
\cite{Kei:2021}.

\textbf{(2)} The IOTA Tangle application in the Internet of Things: Hellani
et al. \cite{Hel:2019}, Gerrits \cite{Ger:2020}, Silvano and Marcelino \cite%
{Sil:2020}, Lee and Sim \cite{Lee:2021}, Wang et al. \cite{Wan:2021}, Zhang
et al. \cite{Zhan:2021} and Igiri et al. \cite{Igi:2022}.

\textbf{(3)} New consensus protocols developed in the DAG-based
blockchain: Danezis et al. \cite{Dan:2018}, Wang et al. \cite{Wan:2019a},
Cui et al. \cite{Cui:2019}, Zhang et al. \cite{Zhan:2020}, Tian et al. \cite%
{Tia:2020}, Reddy and Sharma \cite{Red:2021}, Schett and Danezis \cite%
{Sch:2021}, Xiang et al. \cite{Xia:2021}, Nguyen et al. \cite{Ngu:2021},
Zhou et al. \cite{Zhou:2019}, Spiegelman et al. \cite{Spi:2022}, Deng et al. %
\cite{Deng:2022} and M\"{u}ller et al. \cite{Mul:2022}.

\textbf{(4)} Performance analysis of the DAG-based blockchain: Dong et al. %
\cite{Don:2019}, Fan \cite{Fan:2019}, Fan et al. \cite{Fan:2021}, Guo et al. \cite{Guo:2022} and
Penzkofer et al. \cite{Pen:2020}.

\textbf{(5)} Security of the DAG-based blockchain: Conti et al. \cite%
{Con:2022}, Bramas \cite{Bra:2018, Bra:2021}, Shabandri et al. \cite%
{Sha:2019}, Bhandary et al. \cite{Bha:2020}, Li et al. \cite{Li:2020}, Wang
et al. \cite{Wang:2020, Wan:2020}, Madenouei \cite{Mad:2020}, Brighente et
al. \cite{Bri:2021} and Fan et al. \cite{Fan:2021}. In addition, the
parasite chain attacks are discussed by Staupe \cite{Sta:2017}, Cullen et
al. \cite{Cul:2019} and Penzkofer et al. \cite{Pen:2020}.

\textbf{(6)} Privacy of the DAG-based blockchain: Shabandri et al. \cite%
{Sha:2019}.

\textbf{(7)} Scalability of the DAG-based blockchain: Chen et al. \cite%
{Che:2018}, Wang \cite{Wan:2019}, Madenouei \cite{Mad:2020} and Fan et al. %
\cite{Fan:2021}.

\textbf{(8)} Throughput of the DAG-based blockchain: Zhang et al. \cite%
{Zhan:2020}, Madenouei \cite{Mad:2020}, Brighente et al. \cite{Bri:2021} and
Fan et al. \cite{Fan:2021}.

\textbf{(9)} Stability of the DAG-based blockchain: Bramas \cite{Bra:2018}
and Ferraro et al. \cite{Fer:2019}.

\textbf{(10)} Control of the DAG-based blockchain: Ferraro \cite{Fer:2022},
Vigneri et al. \cite{Vig:2020}, Nakanishi \cite{Nak:2020} and Gupta and
Krishnamurthy \cite{Gup:2022}.

\textbf{(11)} Physical structure of the DAG-based blockchain: A serialized
bolckDAG by Gupta and Janakiram \cite{Gup:2019} and a mixed
Tangle-Blockchain architecture by Hassine et al. \cite{Has:2020}.

\textbf{(12)} Data analysis of the DAG-based Blockchain: Guo et al. \cite%
{Guo:2020}, Gangwani et al. \cite{Gan:2021} , Penzkofer et al. \cite{Pen:2021}
and Silvano and Marcelino \cite{Sil:2020}.

\textbf{(13)} Comparison among DAG, PoW, PoS and BFT: Pervez et al. \cite%
{Per:2018}, Anwar \cite{Anw:2019}, Cao et al. \cite{Cao:2020}, Gerrits \cite%
{Ger:2020}, Khrais \cite{Khr:2020} and Danezis et al. \cite{Dan:2022}.

\textbf{(14)} Useful features of the DAG-based blockchain: Ferraro et al. %
\cite{Fer:2018}, Zou et al. \cite{Zou:2018}, Cao et al. \cite{Cao:2019},
Watanabe et al. \cite{Wat:2019}, Wang et al. \cite{Wan:2019a}, Fan et al. %
\cite{Fan:2019a}, Zhao and Yu \cite{Zhao:2019}, Zhou et al. \cite{Zhou:2019}%
, Birmpas et al. \cite{Bir:2020}, Yin et al. \cite{Yin:2020}, Ding and Sato %
\cite{Din:2020}, Gao et al. \cite{Gao:2020}, Tian et al. \cite{Tia:2020},
Zhang et al. \cite{Zhan:2021, Zhan:2020}, Hellani et al. \cite{Hel:2021},
Wang et al. \cite{Wan:2021}, Jay et al. \cite{Jay:2021}, Nguyen et al. \cite%
{Ngu:2021}, Liao et al. \cite{Lia:2022}, M\"{u}ller et al. \cite{Mul:2022}.

To further improve the performance of the IOTA Tangle (including the above 14 aspects), it is a key to provide a performance analysis
of DAG-based blockchain systems. However, it is interesting but difficult
and challenging to set up a mathematical model for analyzing performance
of the DAG-based blockchain systems. To this end, so far, there have been two
different research classes as follows: \textit{(a) Analytical models} by
Dong et al. \cite{Don:2019}, Fan \cite{Fan:2019}, Park and Kim \cite%
{Park:2019}, Park et al. \cite{Par:2019}, Li et al. \cite{Li:2020}, Cao et
al. \cite{Cao:2020}, Birmpaset al. \cite{Bir:2020} and Fan et al. \cite%
{Fan:2021}; and \textit{(b) simulation models} by Zander et al. \cite%
{Zan:2018}, Bottone et al. \cite{Bot:2018}, Lathif et al. \cite{Lat:2018},
Madenouei \cite{Mad:2020} and Wang et al. \cite{Wan:2020}.

Different from those above works in the literature, this paper provides a
Markov process theory in the study of DAG-based blockchain. By using our
Markov process, for a DAG-based blockchain system, we can provide its
stability analysis, performance evaluation, and optimal control. Therefore,
our Markov process theory can be regarded as a key advance in the study of
DAG-based blockchain.

To show how to set up a Markov process, it is necessary and useful to
explain the IOTA Tangle. The transactions are stored in a DAG,
referred to as Tangle. IOTA uses the PoW and transaction validation
to set up the DAG-based blockchain. Note that the IOTA transactions are
atomic in the Tangle, thus one block corresponds to a transaction. The IOTA
transactions are connected via directed edges to other transactions, where a
directed edge means that a transaction (head) is approved by another (tail).
To issue a transaction and store it in a DAG permanently, the issuer must
approve two other transactions. Any transaction that has not yet been
approved is called a tip. The issuer uses a tip selection algorithm to
choose which tips to approve and uses the PoW to ensure the
validity of such an approval. Based on this, the main purposes of this paper
are to discuss the random behavior of these tips in the DAG and to set up a
Markov process of the tip number.

A few studies close to our work is to apply the simulation models by Ku%
\'{s}mierz \cite{Kus:2017}, Ku\'{s}mierz and Gal \cite{Kus:2018a}, Ku\'{s}%
mierz et al. \cite{ Kus:2019, Kus:2018}, Zanderet at al. \cite{Zan:2018},
Bottone et al. \cite{Bot:2018}, Lathif et al. \cite{Lat:2018}, Chafjiri and
Esfahani \cite{Cha:2019}, Zhao and Yu \cite{Zhao:2019}, Gardner et al. \cite%
{Gar:2020}, Cullen et al. \cite{Cul:2019}, Wang et al. \cite{Wang:2020,
Wan:2021a} and Pere\v{s}\'{\i}ni et al. \cite{Per:2021}. By comparing our
Markov process with those simulation models, we find that our Markov process
can provide more accurate and detailed analysis in the study of DAG-based
blockchain. When using our Markov process, the (tip) random behavior of the
IOTA Tangle can clearly be understood from the network structure of Tip
connections in a time order. In the DAG-based blockchain, each transaction
will have a copy of the stored information; while the network latency causes
inconsistency in the content of the transactions' copies. Thus, this
inconsistency can cause a phenomenon in a deterministic sense that a tip
will be selected multiple times. However, our Markov process will not cause
such a phenomenon that a tip can be selected multiple times. Therefore, this
leads to a great simplification of our Markov model for analyzing the DAG-based
blockchain with IOTA Tangle.

Based on the above analysis, the main contributions of this paper are listed
as follows:

\textbf{(a)} We first describe a simple Markov model for the DAG-based
blockchain with IOTA Tangle by means of two layers of tips and internal
tips' impatient connection behavior. Then we set up a continuous-time Markov
process to analyze the DAG-based blockchain system, and show that the Markov
process is a level-dependent QBD process. Based on this, we prove that this
QBD process must be irreducible and positive recurrent, and can provide
performance analysis of the DAG blockchain system by using the stationary
probability vector of the QBD process.

\textbf{(b) }We propose a new effective method for analyzing the confirmation
time of any arriving internal tip at this system by means of the first passage times
and the PH distributions. Further, we compute the average confirmation time of
any arriving internal tip by means of the RG-factorizations. Crucially, for the complicated DAG-based blockchain system, this paper finds a new effective method to deal with two key performance measures: Throughput and confirmation time.

\textbf{(c) }We use numerical examples to check the validity of our
theoretical results, and to discuss how the performance measures of the
DAG-based blockchain with IOTA Tangle depend on some key parameters of this
system.

Therefore, we believe that the methodology and results developed in this paper
can be applicable to more general DAG-based blockchain systems such that a series of promising
research can be developed potentially.

The remainder of this paper is organized as follows. Section \ref{sec:model} provides a
detailed model description for the DAG-based blockchain system. Section \ref{sec:process}
shows that the DAG-based blockchain system is related to a level-dependent
QBD process, and proves that the DAG-based blockchain system must be positive recurrent. Section \ref{sec:performance} provides performance
analysis of the DAG-based blockchain system. Section \ref{sec:confirmation} proposes a new effective
method for analyzing the confirmation time of any arriving internal tip and computes the average confirmation time by means of
the RG-factorizations. Section \ref{sec:numerical} uses numerical examples to discuss how
performance measures of the DAG-based blockchain system depend on some key
parameters of this system. Section \ref{sec:remarks} gives some concluding remarks.

\section{Model Descriptions}\label{sec:model}
In this section, we provide a model description for a DAG-based blockchain
system and its associated network growth process. Also, we give mathematical
notations, random factors, and necessary parameters.

In a DAG-based blockchain system, its network growth process depends on
the arrivals of new transactions and the increase of confirmed transactions through using a connection algorithm in the
IOTA Tangle. Here, the confirmed transactions are called network nodes (or blocks). To understand the IOTA Tangle easily, the network growth process of the
DAG-based blockchain is depicted in Fig. 1 with some physical
interpretations.

\begin{figure}[h]
\centering
\includegraphics[width=10cm]{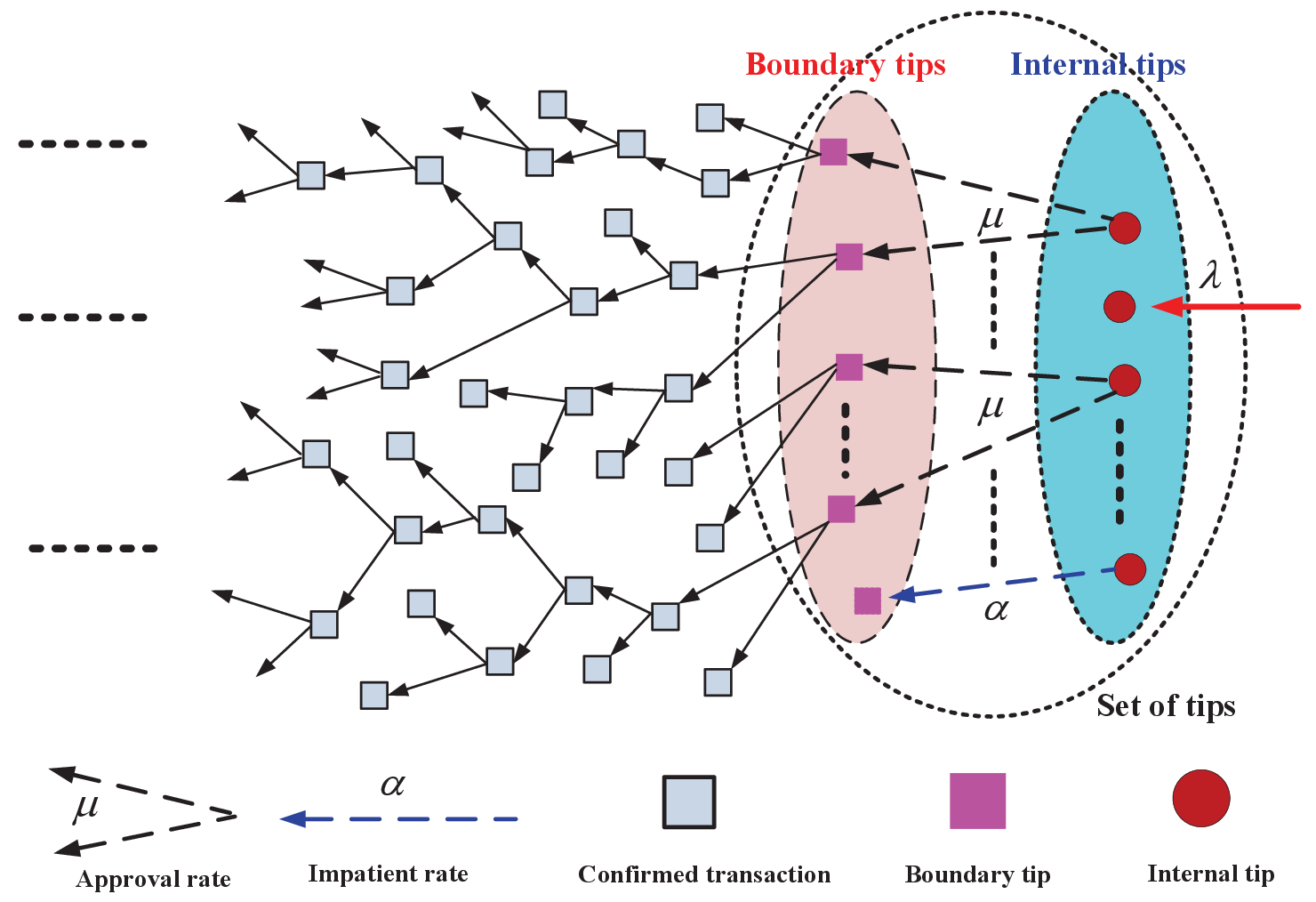}
\caption{Network growth process of DAG-based blockchain system.}
\label{Fig-1}
\end{figure}

By using Fig. 1, we first introduce a key concept: Tips, which are
necessary and useful in our next study. If any transaction is not approved
by another yet, then the transaction is called a tip. In this case, it is
seen from Fig. 1 that all the tips are divided into two different
categories: The boundary tips and the internal tips. Note that there is a
key evolutive characteristic between a boundary tip and an internal tip. If
an internal tip synchronously approves two boundary tips, then the two
boundary tips become two confirmed transactions, so that the two boundary
tips immediately become the network nodes (or blocks) and
leave the set of tips. At the same time, the internal tip immediately
becomes a boundary tip and enters the set of boundary tips.

In this paper, we introduce a key impatient behavior of internal tips. To
prevent the set of internal tips from becoming bigger and bigger such that
the system keeps a highly dynamic nature, we assume that every one of the
internal tips has an impatient behavior. Let $X$ be a nonnegative random
variable. If the waiting time of an internal tip is not less than $X$, then
the internal tip immediately becomes a boundary tip. It is worthwhile to
note that the impatient behavior of internal tips makes that each tip can
leave the set of tips as soon as possible.

By using Fig. 1, we can provide a detailed model description of the DAG-based
blockchain system and its associated network growth process as follows:

\textbf{(1) }For the convenience of discussion, we assume that the capacity
of internal tips is finite such that $S(t)\leq M$, where $M$ is a finite
positive integer. Note that there exists at least one genesis block (i.e.,
an initial transaction) in the set of boundary tips. Thus, we have $1\leq S(t)\leq M$
at any time (see Lemma 1 and its proof in the next section). At the same time, we assume that the capacity of boundary tips
is infinite.

\textbf{(2) Arrivals of new transactions:} Some accounts submit new
transactions into the DAG-based blockchain system according to a Poisson
process with arrival rate $\lambda>0$.

\textbf{(3) Connection of each internal tip with two boundary tips:} With
the same probability, each internal tip in the set of internal tips can
synchronously connect with any two different boundary tips in the set of boundary tips and their connection time is exponential with connection rate $\mu>0$.

In fact, such a connection is an approval, that is, the two boundary tips
are approved by the internal tip. This means that the two boundary tips
become two confirmed transactions so that the two boundary tips immediately
become the network nodes (or blocks) and leave the set of
tips. At the same time, the internal tip immediately becomes a boundary tip
and enters the set of boundary tips.

\textbf{(4) Impatient behavior of each internal tip:} To enable the tips to
enter the network nodes (or blocks), we assume that
each internal tip has an impatient behavior. If the waiting time of the
internal tip exceeds a random impatient time, then this internal tip
directly becomes a boundary tip and enters the set of boundary tips. We
assume that the impatient time of each internal tip is exponential with
inpatient rate $\alpha>0$.

\textbf{(5) Independence:} We assume that all the random variables defined
above are independent of each other.

\begin{Rem}
In the DAG-based blockchain system, we introduce the impatient time of each
internal tip such that the internal tips can be accelerated to become
network nodes. In this case, the throughput of the DAG-based blockchain
system is improved greatly. At the same time, the impatient behavior can
also prevent some tips from staying for too much time in the set of tips.
\end{Rem}

\begin{Rem}
In the DAG-based blockchain system, the purpose of this paper is to
concentrate on the dynamic characteristics of tips changing from the internal tips, to the boundary tips, and finally, to the
network nodes. While introducing some random factors, this paper may be the
first to apply the Markov process theory to the study of DAG-based
blockchain.
\end{Rem}

\section{A Level-Dependent QBD Process}\label{sec:process}
In this section, we consider the network growth process of the DAG-based
blockchain system by means of a continuous-time level-dependent QBD process,
and obtain a stability condition of the DAG-based blockchain system.

Let $Q(t)$ and $S(t)$ be the numbers of the internal tips and the boundary
tips in the DAG-based blockchain system at time $t\geq0$, respectively.
Note that once an internal tip completes the connection (or approval) with
two boundary tips, the two boundary tips become two network nodes so that
both of them leave the set of tips immediately; while the internal
tip becomes a new boundary tip in the set of boundary tips. In this case,
the tip number in the set of boundary tips will be reduced by one. If a new
transaction arrives at the set of internal tips, the tip number in the set
of internal tips will be increased by one. If the internal tip becomes a new
boundary due to the impatient behavior of internal tips, then the tip number
in the set of internal tips will be decreased by one, while the tip number
in the set of boundary tips will be increased by one.

\begin{Lem}
For any time $t\geq0$, $1\leq S(t)\leq M$.
\end{Lem}

\textbf{Proof: }$S(t)\leq M$ comes from Assumption (1) in Section \ref{sec:model}: The
capacity of internal tips is finite. Thus, in what follows we only prove
that $S(t)\geq1$ by induction.

Firstly, for $n=0$ and $t_{0}=0$, $S(0)\geq1$ follows the fact that there
exists at least one genesis block in the set of boundary tips.

Let $t_{1}$ be the epoch that the Markov process $\left\{ S(t):t\geq
0\right\} $ has the first state jumping when it begins at time $0$. Then $%
S(t)=S(0)\geq1$ for $0\leq t<t_{1}$; while $S(t_{1})>S(0)\geq1$ is due to
the fact that a new boundary tip comes from an impatient internal tip, and $
S(t_{1})\geq1$ is due to the fact that once an internal tip completes the connection (or approval) with two
boundary tips, the two boundary tips become two network nodes so that both
of them leave the set of boundary tips immediately, and the internal tip
becomes a new boundary tip in the set of boundary tips.

Secondly, in the cases: $n=1$ to $k-1$, this result holds. Now, we show
that for $n=k$, this result also holds. To this end, we write

\begin{equation*}
t_{k}>t_{k-1}>t_{k-2}>\cdots>t_{1}>t_{0}=0
\end{equation*}
and
\begin{equation*}
S(t_{k-1})>1,S(t_{k-2})>1,\ldots,S(t_{1})>1,S(t_{0})\geq1,
\end{equation*}
where $t_{k}$ be the epoch that the Markov process $\left\{ S(t):t\geq
0\right\} $ has the first state jumping when it begins at time $t_{k-1}$.

Now, we observe this result at time $t_{k}$. Note that $S(t)=S(0)\geq1$ for $%
t_{k-1}\leq t<t_{k}$; while $S(t_{k})>S(t_{k-1})\geq1$ is due to the fact
that a new boundary tip comes from an impatient internal tip, and $%
S(t_{k})\geq1$ is due to the fact that once an internal tip completes the connection (or approval) with two
boundary tips, the two boundary tips become two network nodes so that both
of them leave the set of boundary tips immediately, and the internal tip
becomes a new boundary tip in the set of boundary tips. Therefore, this result can hold at time $t_{k}$. By induction, we show that this result can
hold at time $t\geq0$, i.e., $S(t)\geq1$ for time $t\geq0$.
This completes the proof. $\square$

It is easy to see that $\left\{ \left( Q(t),S(t):t\geq0\right) \right\} $ is
a two-dimensional continuous-time Markov process on a state space, given by%
\begin{equation*}
\Omega=\left\{ \left( n,m\right) :n\geq0,1\leq m\leq M\right\} .
\end{equation*}

Based on this, the state transition relations of this two-dimensional Markov
process $\left\{ \left( Q(t),S(t):t\geq0\right) \right\} $ are depicted in
Fig. 2.

\begin{figure}[ptbh]
\centering \includegraphics[width=10cm]{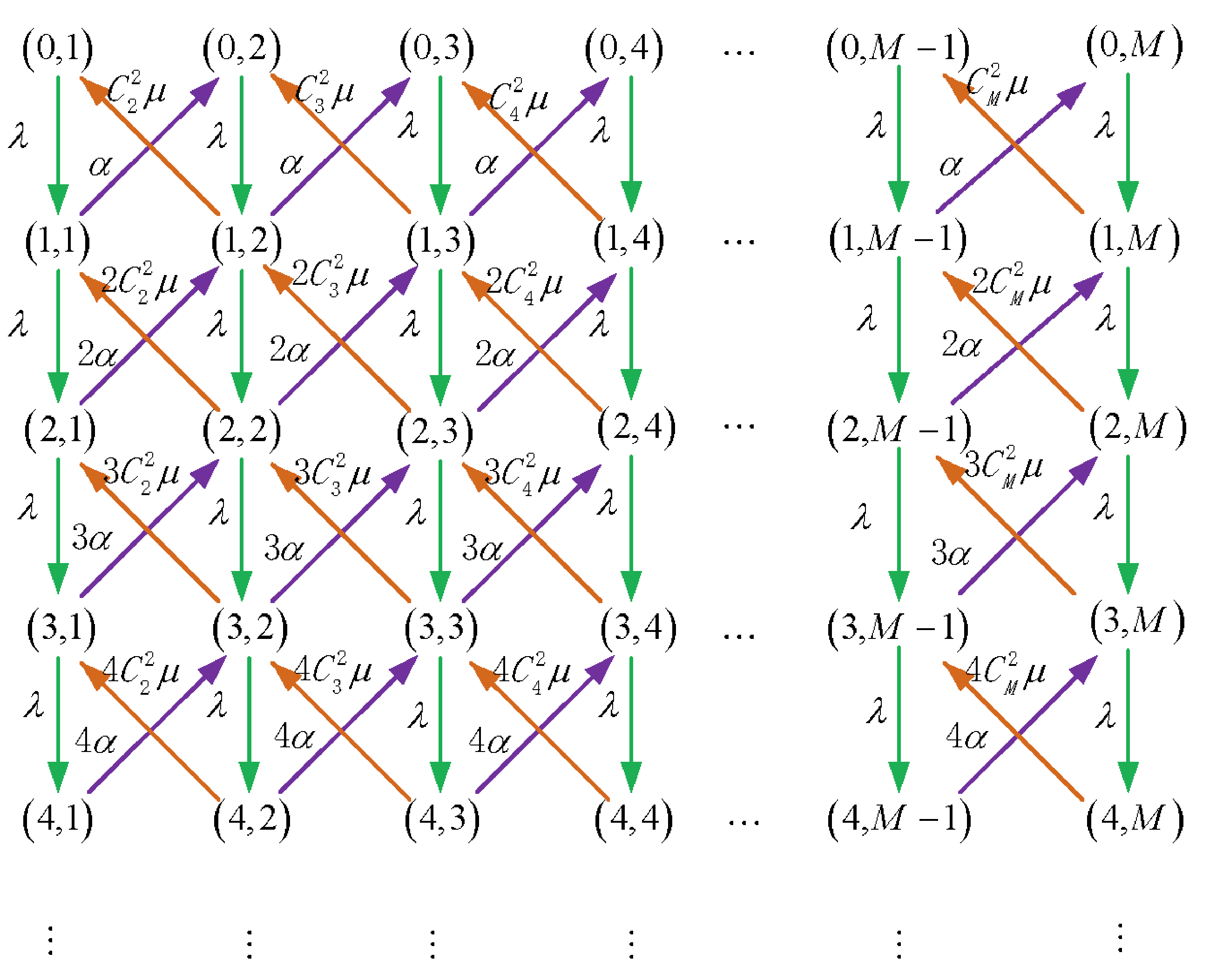}
\caption{The state transition relations of the two-dimensional Markov
process }
\label{figure:fig-2}
\end{figure}

It is easy to see from Fig. 2 that the Markov process $\left\{ \left(
Q(t),S(t):t\geq 0\right) \right\} $ is a continuous-time level-dependent QBD
process whose state space is rewritten as
\begin{equation*}
\Omega =\bigcup\limits_{k=0}^{\infty }\text{Level }k{,}
\end{equation*}%
where
\begin{equation*}
\text{Level }k=\left\{ \left( k,m\right) :1\leq m\leq M\right\} .
\end{equation*}%
Therefore, the Markov process $\left\{ \left( Q(t),S(t):t\geq 0\right)
\right\} $ is a level-dependent QBD process whose infinitesimal generator is
given by
\begin{equation*}
Q=\left( {%
\begin{array}{cccccc}
{{A_{0,0}}} & {{A_{0,1}}} &  &  &  &    \\
{{A_{1,0}}} & {{A_{1,1}}} & {{A_{1,2}}}   &  &    \\
& \ddots & \ddots & \ddots &  &   \\
&  & {{A_{k,k-1}}} & {{A_{k,k}}} & {{A_{k,k+1}}} &  \\
&  &  & \ddots & \ddots & \ddots%
\end{array}%
}\right) ,
\end{equation*}%
where%
\begin{equation*}
{A_{0,0}}=\left(
\begin{array}{ccc}
{-\lambda } &  &    \\
  & \ddots &  \\
  &  & {-\lambda }%
\end{array}%
\right) ,\ {A_{0,1}}=\left(
\begin{array}{ccc}
\lambda &  &    \\
  & \ddots &  \\
 &  & \lambda%
\end{array}%
\right) ;
\end{equation*}%

for $k\geq 1$,
\begin{equation*}
{A_{k,k-1}}=\left(
\begin{array}{ccccc}
0 & k{\alpha } &  &  &  \\
k{C_{2}^{2}\mu } & 0 & k{\alpha } &  &  \\
& \ddots & \ddots & \ddots &  \\
&  & k{C_{M-1}^{2}\mu } & 0 & k{\alpha } \\
&  &  & k{C_{M}^{2}\mu } & 0%
\end{array}%
\right) ,
\end{equation*}
\begin{equation*}
{A_{k,k+1}}=\left(
\begin{array}{ccc}
\lambda &  &   \\
 & \ddots &  \\
   &  & \lambda%
\end{array}%
\right) ,
\end{equation*}

\begin{equation*}
{A_{k,k}}=\left( {%
\begin{array}{ccccc}
-{(k\alpha +\lambda )} &  &  &  &  \\
& {-\left( {kC_{2}^{2}\mu +k\alpha +\lambda }\right) } &  &  &  \\
&  & \ddots &  &  \\
&  &  & {-\left( {kC_{M-1}^{2}\mu +k\alpha +\lambda }\right) } &  \\
&  &  &  & {-\left( {kC_{M}^{2}\mu +\lambda }\right) }%
\end{array}%
}\right) .
\end{equation*}%

In what follows we apply the mean drift method (e.g., see Chapter 3 in Li %
\cite{Li:2010}) to study the stability of the level-dependent QBD process $Q$
corresponding to the DAG-based blockchain system.

The following theorem provides a necessary and sufficient condition under
which the level-dependent QBD $Q$ must be irreducible and positive recurrent.

\begin{The}
The level-dependent QBD $Q$ must be irreducible and positive recurrent. Thus, the DAG-based blockchain system is positive recurrent.
\end{The}

\textbf{Proof: }The irreducibility of this QBD process $Q$ can be directly checked from Fig. 2. Therefore, in what follows we only need to prove that
it is positive recurrent.

Note that the QBD process $Q$ is irreducible, level-dependent and each of
its levels has finite states. Thus, we apply the mean drift method to compare
the upward mean drift rate with the downward mean drift rate on Level $k$
for a larger positive integer $k\geq1$.

For a larger positive integer $k\geq 1$, we have
\begin{align*}
{\mathbf{A}_{k}}& ={A_{k,k-1}}+{A_{k,k}}+{A_{k,k+1}} \\
& =\left(
\begin{array}{ccccc}
{-k\alpha } & {k\alpha } &  &  &  \\
{kC_{2}^{2}\mu } & {-\left( {kC_{2}^{2}\mu +k\alpha }\right) } & {k\alpha }
&  &  \\
& \ddots & \ddots & \ddots &  \\
&  & {kC_{M-1}^{2}\mu } & {-\left( {kC_{M-1}^{2}\mu +k\alpha }\right) } & {%
k\alpha } \\
&  &  & {kC_{M}^{2}\mu } & -{kC_{M}^{2}\mu }%
\end{array}%
\right) .
\end{align*}
It is easy to see that the Markov process $\mathbf{A}_{k}$ is irreducible,
aperiodic and finite states. Thus, the Markov process $\mathbf{A}_{k}$ must
be positive recurrent.

Let $\beta^{(k)}=(\beta_{1}^{(k)},\beta_{2}^{(k)},...,\beta_{M}^{(k)})$ be
the stationary probability vector of the Markov process $\mathbf{A}_{k}$.
Then $\beta^{(k)}\mathbf{A}_{k}=0$ and $\beta^{(k)}e=1$, where $e$ is a
column vector of size $M$ whose components are all ones. Now, we solve the
system of linear equations: $\beta^{(k)}\mathbf{A}_{k}=0$ and $%
\beta^{(k)}e=1 $, that is,
\begin{equation}
\left\{
\begin{array}{l}
-\alpha\beta_{1}^{\left( k\right) }+\mu\beta_{2}^{\left( k\right) }=0, \\
\alpha\beta_{i-1}^{\left( k\right) }-\left( {C_{i}^{2}\mu+\alpha}\right)
\beta_{i}^{\left( k\right) }+C_{i+1}^{2}\mu\beta_{i+1}^{\left( k\right) }=0{,%
}  \quad\quad 2\leq i\leq M-1, \\
\alpha\beta_{M-1}^{\left( k\right) }-\mu C_{M}^{2}\beta_{M}^{\left( k\right)
}=0.%
\end{array}
\right. \   \label{equa-1}
\end{equation}
We obtain
\begin{equation*}
\beta_{2}^{\left( k\right) }=\frac{\alpha}{\mu}\beta_{1}^{\left( k\right) }=%
\frac{1}{{C_{2}^{2}}}\left( {\frac{\alpha}{\mu}}\right) \beta_{1}^{\left(
k\right) },\
\end{equation*}%
\begin{equation*}
\beta_{3}^{\left( k\right) }=\frac{1}{3}{\left( {\frac{\alpha}{\mu}}\right)
^{2}}\beta_{1}^{\left( k\right) }=\frac{1}{{C_{2}^{2}C_{3}^{2}}}{\left( {%
\frac{\alpha}{\mu}}\right) ^{2}}\beta_{1}^{\left( k\right) },\
\end{equation*}%
\begin{equation*}
\beta_{4}^{\left( k\right) }=\frac{1}{{18}}{\left( {\frac{\alpha}{\mu}}%
\right) ^{3}}\beta_{1}^{\left( k\right) }=\frac{1}{{%
C_{2}^{2}C_{3}^{2}C_{4}^{2}}}{\left( {\frac{\alpha}{\mu}}\right) ^{3}}\beta_{1}^{\left( k\right) },
\end{equation*}
for ${5}\leq s\leq M$
\begin{equation*}
\beta_{s}^{\left( k\right) }=\frac{1}{{\prod\limits_{i=2}^{s}{C_{i}^{2}}}}{%
\left( {\frac{\alpha}{\mu}}\right) ^{s-1}}\beta_{1}^{\left( k\right) }{.}\
\end{equation*}
By using $\beta^{(k)}e=1$, we obtain
\begin{equation*}
\beta_{1}^{\left( k\right) }+\frac{1}{{C_{2}^{2}}}\left( {\frac{\alpha}{\mu}}%
\right) \beta_{1}^{\left( k\right) }+\cdots+\frac{1}{{\prod \limits_{i=2}^{M}%
{C_{i}^{2}}}}{\left( {\frac{\alpha}{\mu}}\right) ^{M-1}}\beta_{1}^{\left(
k\right) }=1,
\end{equation*}
this gives%
\begin{equation*}
\beta_{1}^{\left( k\right) }=\frac{1}{1{+\sum\limits_{j=2}^{M}{\frac{1}{{%
\prod\limits_{i=2}^{j}{C_{i}^{2}}}}}{{\left( {\frac{\alpha}{\mu}}\right) }%
^{j-1}}}}\ .
\end{equation*}
Thus the stationary probability vector of the Markov process $\mathbf{A}_{k}$
is given by%
\begin{align*}
\beta_{1}^{\left( k\right) } & =\frac{1}{1{+\sum\limits_{j=2}^{M}{\frac{1}{{%
\prod\limits_{i=2}^{j}{C_{i}^{2}}}}}{{\left( {\frac{\alpha}{\mu}}\right) }%
^{j-1}}}}, \\
\beta_{s}^{\left( k\right) } & =\frac{\frac{1}{{\prod\limits_{i=2}^{s}{%
C_{i}^{2}}}}{\left( {\frac{\alpha}{\mu}}\right) ^{s-1}}}{1{+\sum
\limits_{j=2}^{M}{\frac{1}{{\prod\limits_{i=2}^{j}{C_{i}^{2}}}}}{{\left( {%
\frac{\alpha}{\mu}}\right) }^{j-1}}}},\text{ \ }{2}\leq s\leq M.
\end{align*}
Clearly, the stationary probability vector $\beta^{(k)}$ is independent of
the number $k$.

Now, we calculate the upward and downward mean drift rates on Level $k$. It
is easy to check that the upward mean drift rate, from Level $k$ to Level $%
k+1$, is given by
\begin{equation*}
{\beta^{\left( k\right) }}{A_{k,k+1}}e={\beta^{\left( k\right) }}{\left( {%
\lambda,\lambda,\ldots,\lambda}\right) ^{T}}=\lambda{\beta^{\left( k\right) }%
}e=\lambda.
\end{equation*}
Similarly, the downward mean drift rate, from Level $k$ to Level $k-1$, is
given by%
\begin{align*}
{\beta ^{\left( k\right) }}{A_{k,k-1}}e =&{\beta ^{\left( k\right) }}({{%
k\alpha ,k{ C_{2}^{2}\mu+\alpha ,\ldots ,}}} {{k\left( {C_{M-1}^{2}\mu +\alpha }\right) ,C_{M}^{2}k\mu )}^{T}} \\
=&k\alpha \left( {\beta _{1}^{\left( k\right) }+\beta _{2}^{\left(
k\right) }+\beta _{3}^{\left( k\right) }+\cdots +\beta _{M-1}^{\left(
k\right) }}\right)  \\
&+k\mu \left( {C_{2}^{2}\beta _{2}^{\left( k\right) }+C_{3}^{2}\beta _{3}^{\left( k\right) }+\cdots +C_{M}^{2}\beta _{M}^{\left( k\right) }}\right)
\\
=&k\alpha \left( 1-{\beta _{M}^{\left( k\right) }}\right) +k\mu {%
\sum\limits_{j=2}^{M}C_{j}^{2}\beta _{j}^{\left( k\right) }}.
\end{align*}
Thus, we get
\begin{equation*}
\begin{aligned}
{\beta^{\left( k\right) }}{A_{k,k-1}}e-{\beta^{\left( k\right) }}{A_{k,k+1}}%
e=&k\alpha\left( {1-\beta_{M}^{\left( k\right) }}\right)  +k\mu{%
\sum\limits_{j=2}^{M}C_{j}^{2}\beta_{j}^{\left( k\right) }}-\lambda.
\end{aligned}
\end{equation*}

Note that $\lambda$, $\alpha(1-\beta_{M}^{(k)})$ and $\mu{%
\sum\limits_{j=2}^{M}C_{j}^{2}\beta_{j}^{\left( k\right) }}$ are positive
numbers and they are independent of the number $k$. Thus, it is easy to see
that as $k\rightarrow+\infty$,
\begin{equation*}
\beta^{(k)}A_{k,k-1}e-\beta^{(k)}A_{k,k+1}e\rightarrow+\infty.
\end{equation*}
In this case, there always exists a larger positive integer $K$ such that
for $l\geq K$,
\begin{equation*}
\beta^{(l)}A_{l,l-1}e-\beta^{(l)}A_{l,l+1}e>0.
\end{equation*}
Thus, the downward mean drift rate is bigger than the upward mean drift rate,
that is,
\begin{equation*}
\beta^{(l)}A_{l,l-1}e>\beta^{(l)}A_{l,l+1}e.
\end{equation*}
Therefore, we show that the continuous-time level-dependent QBD process $Q$
is positive recurrent. This completes the proof. $\square$

\section{Performance Analysis}\label{sec:performance}
In this section, we first compute the stationary probability vector of the
level-dependent QBD process by means of the RG-factorizations given in Li %
\cite{Li:2010}. Then we provide performance analysis of the DAG-based
blockchain system.

Let
\begin{equation*}
p_{k,i}\left( t\right) =P\left\{ Q\left( t\right) =k{,}S\left( t\right)
=i\right\} , k\geq0,1\leq i\leq M.
\end{equation*}
Then since the level-dependent QBD process $Q$ is irreducible and positive
recurrent, we have
\begin{equation*}
\pi_{k,i}=\lim_{t\rightarrow\infty}p_{k,i}\left( t\right) .
\end{equation*}
We write%
\begin{equation*}
\pi_{k}=\left( \pi_{k,1},\pi_{k,2}{,\ldots,}\pi_{k,M}\right) ,\text{ \ }%
k\geq0,
\end{equation*}
and
\begin{equation*}
\pi=\left( \pi_{0},\pi_{1},\pi_{2}{,\ldots}\right).
\end{equation*}

Note that the QBD process is level-dependent, we need to use the
RG-factorizations to compute the stationary probability vector. To this end,
we introduce $R$-, $U$- and $G$- measures as follows:
\begin{equation*}
{U_{k}}={A_{k,k}}+{A_{k,k+1}}\left( {-U_{k+1}^{-1}}\right) {A_{k+1,k}},\text{%
{\ }}k\geq0,
\end{equation*}%
\begin{equation*}
{R_{k}}={A_{k,k+1}}\left( {-U_{k+1}^{-1}}\right) ,\text{{\ }}k\geq0,
\end{equation*}%
\begin{equation*}
{G_{k}}=\left( {-U_{k+1}^{-1}}\right) {A_{k+1,k}},\text{{\ }}k\geq1.
\end{equation*}

On the other hand, by using Ramaswami and Taylor \cite{Ram:1996}, the matrix
sequence $\{{{R_{k}},k}\geq0\}$ is the minimum nonnegative solution of the
system of nonlinear matrix equations
\begin{equation*}
{A_{k,k+1}}+{R_{k}}{A_{k+1,k+1}}+{R_{k}}{R_{k+1}}{A_{k+2,k+1}}=0,{k}\geq0,\
\end{equation*}
and the matrix sequence $\{{{G_{k}},k}\geq1\}$ is the minimum nonnegative
solution of the system of nonlinear matrix equations
\begin{equation*}
{A_{k,k+1}}{G_{k+1}}{G_{k}+{A_{k,k}}{G_{k}+}A_{k,k-1}}=0,\text{{\ }}{k}%
\geq1.\
\end{equation*}
Once $\{{{R_{k}},k\geq0}\}$ or $\{{{G_{k}},k}\geq1\}$ is determined, we can
get
\begin{equation*}
{U_{k}}={A_{k,k}}+{R_{k}}{A_{k+1,k}=A_{k,k}}+{A_{k,k+1}}{G_{k+1}},\text{{\ }}%
{k}\geq0.
\end{equation*}

By using the $R$-, $U$- and $G$- measures and Li \cite{Li:2010} or Li and
Cao \cite{Li:2004}, the UL-type RG-factorization of the QBD process is given
by
\begin{equation*}
Q=\left( {I-{R_{U}}}\right) {U_{D}}\left( {I-{G_{L}}}\right) ,
\end{equation*}
where
\begin{equation*}
{U_{D}}=\text{diag}\left( {{U_{0}},{U_{1}},{U_{2}},\ldots}\right) ,
\end{equation*}

\begin{equation*}
{R_{U}}=\left( {%
\begin{array}{ccccc}
0 & {{R_{0}}} &  &  &  \\
& 0 & {{R_{1}}} &  &  \\
&  & 0 & {{R_{2}}} &  \\
&  &  & \ddots & \ddots%
\end{array}
}\right) ,
\end{equation*}
\begin{equation*}
\text{{\ }}{G_{L}}=\left( {%
\begin{array}{ccccc}
0 &  &  &  &  \\
{{G_{1}}} & 0 &  &  &  \\
& {{G_{2}}} & 0 &  &  \\
&  & {{G_{3}}} & 0 &  \\
&  &  & \ddots & \ddots%
\end{array}
}\right) .
\end{equation*}

\begin{The}
The stationary probability vector $\pi$ of the level-dependent QBD process $%
Q $\ is given by
\begin{equation*}
{\pi_{k}}=c{\tilde{\pi}_{k}},\text{{\ }}k\geq0,\
\end{equation*}%
\begin{equation*}
{\tilde{\pi}_{k}}={\tilde{\pi}_{0}}{R_{0}}{R_{1}}\cdots{R_{k-1}},\text{{\ }}%
k\geq1,
\end{equation*}
where $\tilde{\pi}_{0}$ is uniquely determined by the system of linear
equations
\begin{equation*}
{\tilde{\pi}_{0}}\left( {{A_{0,0}}+{R_{0}A_{1,0}}}\right) =0,\
\end{equation*}%
\begin{equation*}
{\tilde{\pi}_{0}}e=1,\
\end{equation*}
and the constant $c$ is given by
\begin{equation*}
c=\frac{1}{{{{\tilde{\pi}}_{0}}e+\sum\limits_{k=1}^{\infty}{{{\tilde{\pi}}%
_{0}}{R_{0}}{R_{1}}\cdots{R_{k-1}}e}}}.
\end{equation*}
\end{The}

Based on the stationary probability vector, we provide some steady-state
performance measures of the DAG-based blockchain system as follows:

(1) The steady-state average number of internal tips in the DAG-based
blockchain system is given by%
\begin{align*}
E\left[ {{N_{A}}}\right] = & \left[
{\left( {{\pi_{1,1}+\pi_{1,2}+\pi_{1,3}+}\cdots+{\pi_{1,M}}}\right)}\right.\\
&+2\left( {{\pi_{2,1}+\pi_{2,2}+\pi_{2,3}+}\cdots+{\pi_{2,M}}}\right)\\
& \left. {+3\left( {{\pi_{3,1}+\pi_{3,2}+\pi_{3,3}+}\cdots+{\pi_{3,M}}}\right)
+\cdots}
\right] \\
= & \sum\limits_{{k}=1}^{\infty}{k\cdot{\pi_{k}}e}=c\sum\limits_{k=1}^{%
\infty}{k\cdot{{\tilde{\pi}}_{k}}e} \\
= & c\sum\limits_{{k}=1}^{\infty}{k{{\tilde{\pi}}_{1}}{R_{1}}}{R_{2}}\cdots{%
R_{k-1}}e.
\end{align*}

(2) The steady-state average number of boundary tips in the DAG-based
blockchain system is given by%
\begin{align*}
E\left[ {{N_{B}}}\right] = & \left[
{\left( {{\pi_{0,1}}+{\pi_{1,1}+\pi_{2,1}+\pi_{3,1}+}\cdots}\right)}\right.\\
&+2\left( {{\pi_{0,2}}+{\pi_{1,2}+\pi_{2,2}+\pi_{3,2}+}\cdots}\right)\\
& \left. {+M\left( {{\pi_{0,M}}+{\pi_{1,M}+\pi_{2,M}+\pi_{3,M}+}\cdots }\right)}
\right] \\
= & \sum\limits_{k=0}^{\infty}{{\pi_{k}}{f_{A}=}}c{{\tilde{\pi}}_{0}}{f_{A}}%
+c\sum\limits_{k=1}^{\infty}{{{\tilde{\pi}}_{0}}{R_{0}}}{R_{1}}\cdots {%
R_{k-1}}{f_{A},}
\end{align*}
where $f_{A}=(1,2,...,M)^{T}$.

(3) The throughput of the DAG-based blockchain system

Note that the throughput of the DAG-based blockchain system is defined as
the number of new network nodes generated per unit of time, which is an important indicator to measure the DAG-based blockchain system.

The following theorem provides expression for throughput of the DAG-based blockchain system.

\begin{The}
If the DAG-based blockchain system is stable, then%
\begin{align*}
\text{TH}= & 2\mu\left[ {\left( {{\pi_{1,2}+C}_{3}^{2}{\pi_{1,3}+C}_{4}^{2}{%
\pi_{1,4}+}\cdots{+C}_{M}^{2}{\pi_{1,M}}}\right) }\right. \\
& +2\left( {{\pi_{2,2}+C}_{3}^{2}{\pi_{2,3}+C}_{4}^{2}{\pi_{2,4}+}\cdots {+C}%
_{M}^{2}{\pi_{2,M}}}\right) \\
& +\left. {3\left( {{\pi_{3,2}+C}_{3}^{2}{\pi_{3,3}+C}_{4}^{2}{\pi_{3,4}+}%
\cdots{+C}_{M}^{2}{\pi_{3,M}}}\right) +\cdots}\right] \\
= & 2\mu\sum\limits_{k=1}^{\infty}k{{\pi_{k}}}{f,}
\end{align*}
where $f=(0,1,C_{3}^{2},...,C_{M-1}^{2},C_{M}^{2})^{T}$.
\end{The}

\textbf{Proof: }From Fig. 2, we observe that all the states $\left(
m,n\right) $ for $m\geq1$ and $n\geq2$ can generate the network nodes (or blocks), seeing those bottom states of the brown arrows. Note
that \text{TH} is the product of $2\mu$ and the state probability sum that can
generate the network nodes. Let $P_\text{TH}$ denote the state probability sum that can
generate the network nodes. Accordingly, we need to compute the state probability sum that can generate the network
nodes.

For state $\left( 1,2\right) $, the state probability that can generate the
network nodes is equal to $C_{2}^{2}\pi_{1,2}$. For state $\left( 1,3\right)
$, the state probability that can generate the network nodes is equal to $%
C_{3}^{2}\pi_{1,3}$. For state $\left( 1,M\right) $, the state probability
that can generate the network nodes is equal to $C_{M}^{2}\pi_{1,M}$.

For state $\left( 2,2\right) $, the state probability that can generate the
network nodes is equal to $2C_{2}^{2}\pi_{2,2}$. For state $\left(
2,3\right) $, the state probability that can generate the network nodes is
equal to $2C_{3}^{2}\pi_{2,3}$. For state $\left( 2,M\right) $, the state
probability that can generate the network nodes is equal to $%
2C_{M}^{2}\pi_{2,M}$.

For state $\left( m,2\right) $, the state probability that can generate the
network nodes is equal to $mC_{2}^{2}\pi_{m,2}$. For state $\left(
m,3\right) $, the state probability that can generate the network nodes is
equal to $mC_{3}^{2}\pi_{m,3}$. For state $\left( m,M\right) $, the state
probability that can generate the network nodes is equal to $%
mC_{M}^{2}\pi_{m,M}$.

Based on analysis of the above special cases, we obtain%
\begin{align*}
P_\text{TH}= & {\left( {{\pi_{1,2}+C}_{3}^{2}{\pi_{1,3}+C}_{4}^{2}{\pi_{1,4}+%
}\cdots{+C}_{M}^{2}{\pi_{1,M}}}\right) } \\
& +2\left( {{\pi_{2,2}+C}_{3}^{2}{\pi_{2,3}+C}_{4}^{2}{\pi_{2,4}+}\cdots {+C}%
_{M}^{2}{\pi_{2,M}}}\right) \\
& +{3\left( {{\pi_{3,2}+C}_{3}^{2}{\pi_{3,3}+C}_{4}^{2}{\pi_{3,4}+}\cdots{+C}%
_{M}^{2}{\pi_{3,M}}}\right) +\cdots} \\
= & \sum\limits_{k=1}^{\infty}k{{\pi_{k}}}{f,}
\end{align*}
Thus, we obtain%
\begin{equation*}
\text{TH}=2\mu\times P_\text{TH}=2\mu\sum\limits_{k=1}^{\infty}k{{\pi_{k}}}{%
f.}
\end{equation*}
This completes the proof. $\square$

\begin{Rem}
Under the complicated structure of the DAG-based blockchain systems, this
paper may be the first one to provide an exact computational formula for the
throughput of the DAG-based blockchain systems.
\end{Rem}

\section{The Confirmation Time of Any Arriving Internal tip}\label{sec:confirmation}
In this section, we provide a new efficient method to compute the average
confirmation time of any arriving internal tip (or arriving transaction) at the DAG-based
blockchain system.

The confirmation time of any arriving internal tip at the DAG-based blockchain system is
a time interval from the arrival epoch of the new transaction to the time
that it finally becomes a network node (i.e. block). Obviously, such a confirmation time
includes two different processes: the internal tip is changed to a boundary
tip, and the boundary tip is changed to a network node.

Let $W_{A}$ be the confirmation time of the arriving internal tip A in the DAG-based
blockchain system. For the convenience of computation, we assume that the
arriving internal tip A arrives at the DAG-based blockchain system at time $0$.

Let $I(t)$ and $J(t)$ denote the number of internal tips excluding the
arriving internal tip A and the number of boundary tips excluding the one generated
from the arriving internal tip A in the DAG-based blockchain system at time $t$,
respectively. Let $\Delta$ be an absorbing state that the arriving internal tip A is
finally changed to a network node.

It is easy to see that $\{(1,I(t);J(t)),(I(t);1,J(t)){:}$ $t\geq{0}\}$ is a
Markov process with the absorbing state $\Delta$. In state $(1,I(t);J(t))$, $%
1$ means the arriving internal tip A; in state $(I(t);1,J(t))$, $1$ means a boundary
tip who is generated from the arriving internal tip A. Based on this, the state space
of the Markov process $\{(1,I(t);J(t)),(I(t);1,J(t)){:}$ $t\geq{0}\}$ is given by%

\begin{equation*}
\Theta=\left\{ \Delta\right\} \cup\left\{ \bigcup\limits_{k=0}^{\infty }%
\text{Level }k\right\} ,
\end{equation*}
where,%
\begin{align*}
\text{Level }k=&\left\{{ \left( 1,k;1\right) ,(k;1,0);\left( 1,k;2\right),(k;1,1);\ldots;}\right.\left.{\left( 1,k;M\right) ,(k;1,M-1)}\right\}.
\end{align*}

\begin{figure}[!t]
\centering
\subfloat[]{\includegraphics[width=5.5cm,angle=90]{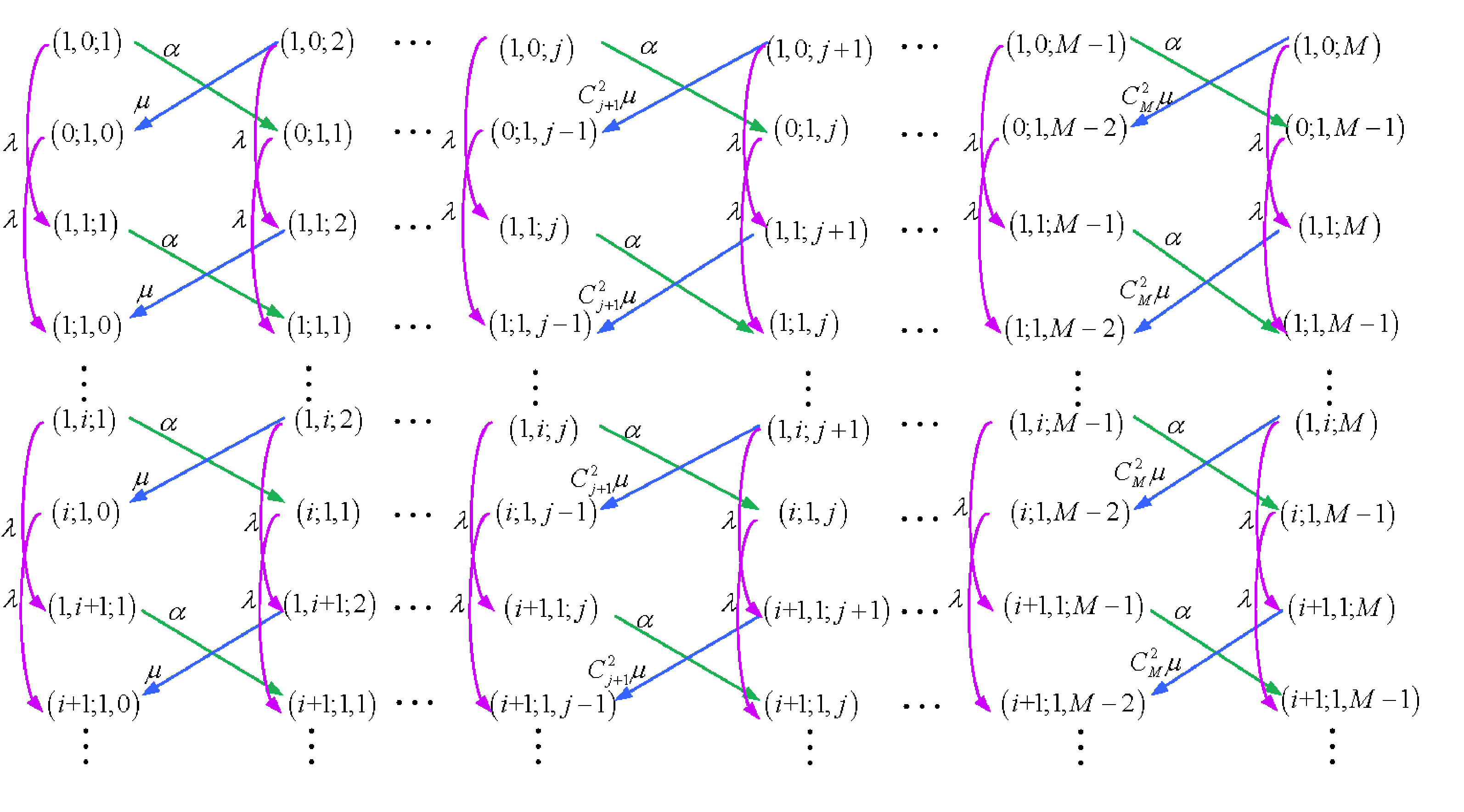}\label{fig-3a}}
\hfil
\subfloat[]{\includegraphics[width=5.5cm,angle=90]{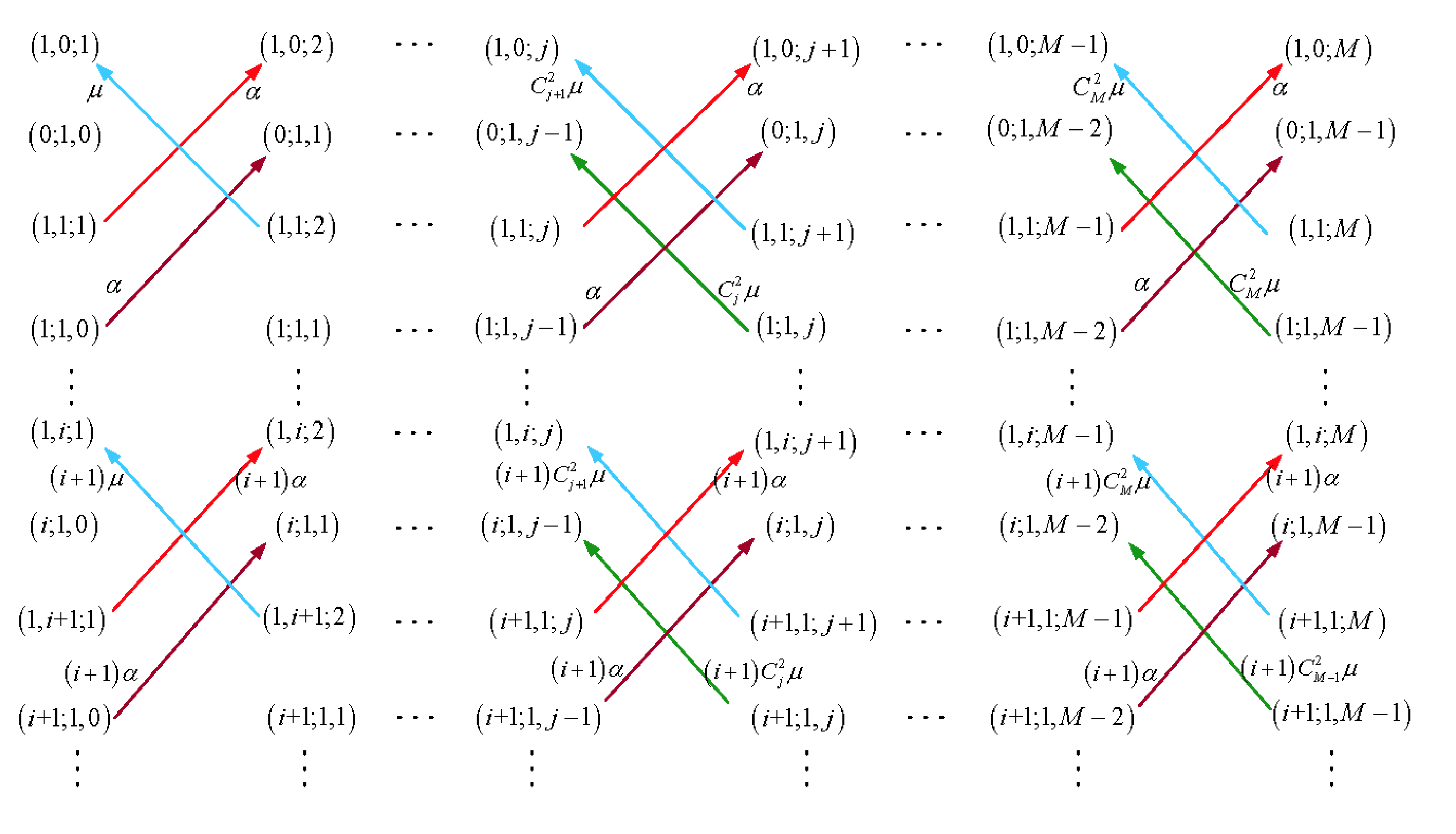}\label{fig-3b}} %
\hfil
\subfloat[]{\includegraphics[width=5.5cm,angle=90]{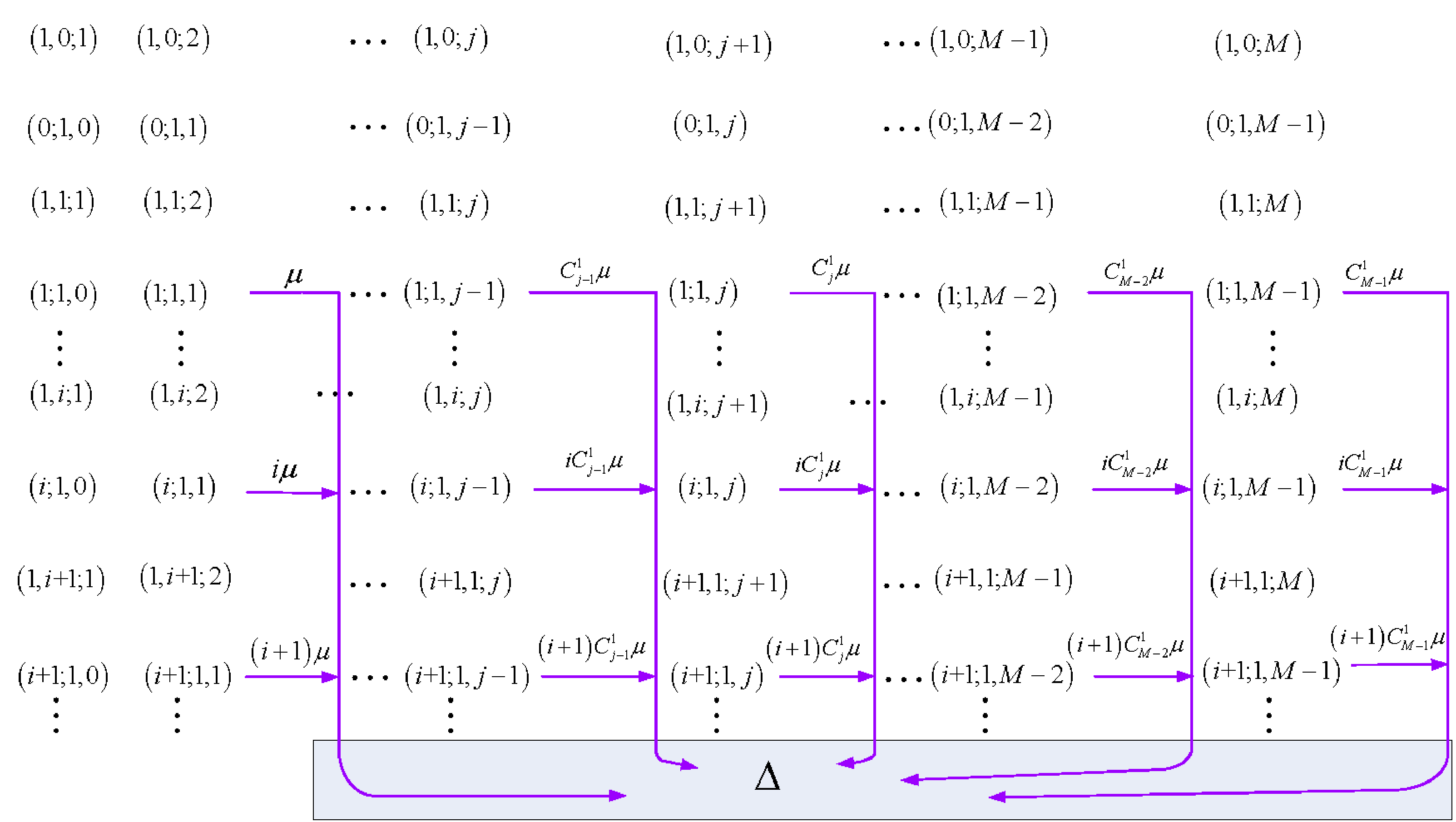}\label{fig-3c}}
\caption{State transition relations of the Markov process.}
\label{fig-3}
\end{figure}

By using the level structure of the state space $\Theta$, Fig. \ref{fig-3a}, \ref{fig-3b}
and \ref{fig-3c} depict the state transition relations of the Markov process $\{(1,I(t);J(t)),(I(t);1,J(t)){:}$ $t\geq{0}\}$.

\begin{Rem}
(i) Fig. \ref{fig-3a} contains three events: (1) Arrivals of internal tips
(i.e., new transactions), (2) the internal tip A is changed to the boundary
tip A due to its impatient behavior, and (3) the internal tip A, connecting (or approving) two boundary tips, is changed to the boundary tip A.

(ii) Fig. \ref{fig-3b} contains two events: (1) The internal tips
excluding the internal tip A are changed to the boundary tips due to their
impatient behavior, and (2) the internal tips, excluding the internal tip A, connecting (or approving) two boundary tips, are changed to the boundary
tips.

(iii) Fig. \ref{fig-3c} contains one event: The boundary tip A is changed
to a network node.
\end{Rem}

From Fig. \ref{fig-3a}, \ref{fig-3b} and \ref{fig-3c}, it is easy to check that the infinitesimal generator of the Markov process $\{(1,I(t);J(t)),(I(t);1,J(t)){:}$ $t\geq {0}\}$ with the absorbing state $\Delta $ is given by
\begin{equation*}
\mathbf{T}=\left(
\begin{array}{cc}
0 & \mathbf{0} \\
T{^{\Delta }} & T%
\end{array}%
\right)
\end{equation*}%
where
\begin{equation*}
T{^{\Delta }}+Te=0,
\end{equation*}

\begin{equation*}
T=\left(
\begin{array}{ccccc}
{{T_{0,0}}} & {{T_{0,1}}} &  &  &  \\
{{T_{1,0}}} & {{T_{1,1}}} & {{T_{1,2}}} &  &  \\
& {{T_{2,1}}} & {{T_{2,2}}} & {{T_{2,3}}} &  \\
&  & \ddots & \ddots & \ddots%
\end{array}%
\right) ,
\end{equation*}

\begin{equation*}
T{^{\Delta }}=\left(
\begin{array}{c}
T{_{0}^{\Delta }} \\
T{_{1}^{\Delta }} \\
T{_{2}^{\Delta }} \\
\vdots%
\end{array}
\right) ,
\end{equation*}

\begin{equation*}
T_{i}^{\Delta }=\left( 0,0;0,{iC_{1}^{1}\mu ;}0,{iC_{2}^{1}\mu ;}0,{%
iC_{3}^{1}\mu ;\ldots ;}0,{iC_{M-1}^{1}\mu }\right) ^{T};
\end{equation*}

\begin{equation*}
T{_{i,i+1}=}\left( {%
\begin{array}{cccc}
\lambda &  &  &  \\
& \lambda &  &  \\
&  & \ddots &  \\
&  &  & \lambda%
\end{array}
}\right) ,\text{ }i\geq0;
\end{equation*}

\begin{equation*}
T{_{0,0}=}\left(
\begin{array}{ccccc}
{{A_{1,1}}} & {{A_{1,2}}} &  &  &  \\
{{A_{2,1}}} & {{A_{2,2}}} & {{A_{2,3}}} &  &  \\
& \ddots & \ddots & \ddots &  \\
&  & {{A_{M-1,M-2}}} & {{A_{M-1,M-1}}} & {{A_{M-1,M}}} \\
&  &  & {{A_{M,M-1}}} & {{A_{M,M}}}%
\end{array}%
\right) ,
\end{equation*}

\begin{equation*}
{A_{1,1}}=\left( {%
\begin{array}{cc}
{-\left( {\alpha+\lambda}\right) } &  \\
& {-\lambda}%
\end{array}
}\right) ,{\ \ A_{1,2}}=\left( {%
\begin{array}{cc}
& \alpha \\
&
\end{array}
}\right) ,
\end{equation*}
for $2\leq l\leq M-1,$

\begin{equation*}
{A_{l,l-1}}=\left( {%
\begin{array}{cc}
& {C_{l}^{2}\mu} \\
&
\end{array}
}\right) ,
{A_{l,l+1}}=\left( {%
\begin{array}{cc}
\alpha &  \\
&
\end{array}
}\right) ,
\end{equation*}

\begin{equation*}
{A_{l,l}}=\left( {%
\begin{array}{cc}
{-\left( {C_{l}^{2}\mu+\alpha+\lambda}\right) } &  \\
& {-\lambda}%
\end{array}
}\right) ,
\end{equation*}

\begin{equation*}
{A_{M,M-1}}=\left( {%
\begin{array}{cc}
& {C_{M}^{2}\mu} \\
&
\end{array}
}\right) ,
{A_{M,M}}=\left( {%
\begin{array}{cc}
{-\left( {C_{M}^{2}\mu+\lambda}\right) } &  \\
& {-\lambda}%
\end{array}
}\right) ;
\end{equation*}
\begin{equation*}
T{_{i,i-1}}=\left(
\begin{array}{ccccc}
0 & {{B_{1,2}}} &  &  &  \\
{{B_{2,1}}} & 0 & {{B_{2,3}}} &  &  \\
& \ddots  & \ddots  & \ddots  &  \\
&  & {{B_{M-1,M-2}}} & 0 & {{B_{M-1,M}}} \\
&  &  & {{B_{M,M-1}}} & 0%
\end{array}
\right) ,
\end{equation*}
for $1\leq l\leq M-1,$
\begin{equation*}
{B_{l,l+1}}=\left( {%
\begin{array}{cc}
{i\alpha } &  \\
& {i\alpha }%
\end{array}%
}\right) ,
{B_{2,1}}=\left( {
\begin{array}{cc}
{i\mu} &  \\
& 0%
\end{array}
}\right) ,
\end{equation*}
for $3\leq l\leq M,$
\begin{equation*}
{B_{l,l-1}}=\left( {
\begin{array}{cc}
{iC_{l}^{2}\mu} &  \\
& {iC_{l-1}^{2}\mu}%
\end{array}
}\right) ;
\end{equation*}
\begin{equation*}
T{_{i,i}}=\left(
\begin{array}{cccccc}
{{C_{1,1}}} & {{C_{1,2}}} &  &  &  &  \\
{{C_{2,1}}} & {{C_{2,2}}} & {{C_{2,3}}} &  &  &  \\
& \ddots & \ddots & \ddots &  &  \\
&  & {{C_{M-2,M-3}}} & {{C_{M-2,M-2}}} & {{C_{M-2,M-1}}} &  \\
&  &  & {{C_{M-1,M-2}}} & {{C_{M-1,M-1}}} &  \\
&  &  &  & {{C_{M,M-1}}} & {{C_{M,M}}}%
\end{array}%
\right) ,i\geq 1,
\end{equation*}
\begin{equation*}
{C_{1,1}}=\left( {
\begin{array}{cc}
{-\left( {\alpha+i\alpha+\lambda}\right) } &  \\
& {-\left( {i\alpha+\lambda}\right) }%
\end{array}
}\right) ,
\end{equation*}

\begin{equation*}
{C_{1,2}}=\left( {
\begin{array}{cc}
& \alpha \\
&
\end{array}
}\right) ,
{C_{2,1}}=\left( {
\begin{array}{cc}
& \mu \\
&
\end{array}
}\right) ,{C_{2,3}}=\left( {%
\begin{array}{cc}
& \alpha \\
&
\end{array}
}\right) ,
\end{equation*}%
\begin{equation*}
{C_{2,2}}=\left( {%
\begin{array}{cc}
{-\left( {\mu+\alpha+i\mu+i\alpha+\lambda}\right) } &  \\
& {-\left( {i\alpha+\lambda+i\mu}\right) }%
\end{array}
}\right) ,
\end{equation*}

\begin{equation*}
{C_{M-1,M-2}}=\left( {%
\begin{array}{cc}
& {C_{M-1}^{2}\mu} \\
&
\end{array}
}\right) ,
{C_{M,M-1}}=\left( {%
\begin{array}{cc}
& {C_{M}^{2}\mu} \\
&
\end{array}
}\right) ,
\end{equation*}

\begin{equation*}
{C_{M-1,M-1}}=\left( {%
\begin{array}{cc}
{-\left( {C_{M-1}^{2}\mu+iC_{M-1}^{2}\mu+i\alpha+\lambda}\right) } &  \\
& {-\left( {iC_{M-2}^{2}\mu+i\alpha+\lambda+iC_{M-2}^{1}\mu}\right) }%
\end{array}
}\right) ,
\end{equation*}
\begin{equation*}
{C_{M,M}}=\left( {%
\begin{array}{cc}
{-\left( {C_{M}^{2}\mu+iC_{M}^{2}\mu+\lambda}\right) } &  \\
& {-\left( {iC_{M-1}^{2}\mu+\lambda+iC_{M-1}^{1}\mu}\right) }%
\end{array}
}\right) ,
\end{equation*}
for $3\leq l\leq M-2,$
\begin{equation*}
{C_{l,l-1}}=\left( {%
\begin{array}{cc}
& {C_{l}^{2}\mu} \\
&
\end{array}
}\right) ,{C_{l,l+1}}=\left( {%
\begin{array}{cc}
& \alpha \\
&
\end{array}
}\right) ,
\end{equation*}
\begin{equation*}
{C_{l,l}}=\left( {%
\begin{array}{cc}
{-\left( {C_{l}^{2}\mu+\alpha+iC_{l}^{2}\mu+i\alpha+\lambda}\right) } &  \\
& {-\left( {iC_{l-1}^{2}\mu+i\alpha+\lambda+iC_{l-1}^{1}\mu}\right) }%
\end{array}
}\right) .
\end{equation*}

Let $(\theta_{\Delta},\theta\mathbf{)}$ be the initial probability vector of
the Markov process $\mathbf{T}$ with an absorbing state $\Delta$ for $%
\theta_{\Delta}=0$, the vector ${\theta}=(0,0,...,0,1,0,...)$ shows that the
sub-Markov process \textbf{$T$ }is at the state $(1,i_{0};j_{0})$ at time $0$.
Therefore, the $\left( 2Mi_{0}+2j_{0}-1\right) $-st element of the vector $%
\theta$ is $1$, and all the other elements are $0$.

The following theorem provides expression for the probability distribution
of the confirmation time $W_{A}$ by means of the first passage times and the
phase-type distributions of infinite sizes.

\begin{The}
The probability distribution of the confirmation time $W_{A}$ is of phase-type
with an irreducible representation $(\theta,T)$, and
\begin{equation*}
{F_{{W_{A}}}}\left( t\right) =P\left\{ W_{A}\leq t\right\} =1-\theta
\exp\left\{ Tt\right\} e,\text{ }t\geq0.
\end{equation*}
Also, the average confirmation time $E\left[ {{W_{A}}}\right]$ is given by
\begin{equation*}
E\left[ {{W_{A}}}\right] =-\theta T_{\max}^{-1}e,
\end{equation*}
where $T_{\max}^{-1}$ is the maximal non-positive inverse of the matrix $T$
of infinite sizes.
\end{The}

\textbf{Proof: }For $i\geq 0,$ $1\leq j\leq M$ and $0\leq n\leq M-1$, we
write
\begin{equation*}
q_{1,i;j}\left( t\right) =P\left\{ I\left( t\right) =i{,}J\left( t\right)
=j\right\} ,
\end{equation*}%
\begin{equation*}
q_{i;1,n}\left( t\right) =P\left\{ I\left( t\right) =i,J\left( t\right)
=n\right\} ,
\end{equation*}%
\begin{equation*}
\begin{aligned}
q_{i}\left( t\right) =&\left( {q_{1,i;1}\left( t\right) ,q_{i;1,0}\left(
t\right) ;q_{1,i;2}\left( t\right) ,q_{i;1,1}\left( t\right) ;\ldots;}\right.
\left.{q_{1,i;M}\left( t\right) ,q_{i;1,M-1}\left( t\right) }\right) ,
\end{aligned}
\end{equation*}%
\begin{equation*}
q\left( t\right) =\left( q_{0}\left( t\right) ,q_{1}\left( t\right)
,q_{2}\left( t\right) ,\ldots \right) .
\end{equation*}%
By using the Chapman-Kolmogorov forward differential equation, we obtain
\begin{equation}
\frac{\mathrm{d}}{{\mathrm{d}t}}q\left( t\right) =q\left( t\right) T,\
\label{equa-2}
\end{equation}%
with the initial conditions
\begin{equation}
q\left( 0\right) =\theta .\   \label{equa-3}
\end{equation}%
It follows from (\ref{equa-2}) and (\ref{equa-3}) that
\begin{equation}
q\left( t\right) =\theta \exp \left\{ Tt\right\} {.}\   \label{equa-4}
\end{equation}%
Note that $q(0)e=1$, it follows from (\ref{equa-4}) that
\begin{equation*}
{F_{W_{A}}}\left( t\right) =P\left\{ W_{A}\leq t\right\} =1-\theta \exp
\left\{ Tt\right\} e,\text{{\ }}t\geq 0.
\end{equation*}

In what follows we compute the average confirmation time $E\left[ W_{A}\right] $
by the Laplace-Stieltjes transform. Let $f(s)$ be the Laplace-Stieltjes
transform of the distribution function ${F_{{W_{A}}}}\left( t\right) $ (or
the random variable $W_{A}$). Then

\begin{equation*}
f\left( s\right) =\int_{0}^{\infty}{{e^{-st}}}\text{\textrm{d}}{F_{{W_{A}}}}%
\left( t\right) =1+\theta\left( sI-T\right) _{\min}^{-1}T^{0},\
\end{equation*}
where $\left( sI-T\right) _{\min}^{-1}$ is the minimal non-negative inverse
of the matrix $sI-T$ of infinite sizes for $s\geq0$. Hence we obtain
\begin{equation*}
\begin{aligned}
E\left[ W_{A}\right] &=-\frac{\mathrm{d}}{{\mathrm{d}s}}f{\left( s\right) _{
\mathrm{|}s=0}}
&=\theta{\left[ \left( sI-T\right) _{\min}^{-2}\right] _{|s=0}}T^{0}
&=-\theta T_{\max}^{-1}e.
\end{aligned}
\end{equation*}
This completes the proof. $\square$

To compute the inverse matrix $T_{\max}^{-1}$ of infinite size, we need to use the
RG-factorizations of the Markov process $T$. To this end, we introduce $R$-, $%
U$- and $G$- measures as follows:
\begin{equation*}
U_{i}=T_{i,i}+T_{i,i+1}\left( {-}U_{i+1}^{-1}\right) T_{i+1,i},\text{ \ }%
i\geq0,
\end{equation*}%
\begin{equation*}
R_{i}=T_{i,i+1}\left( -U_{i+1}^{-1}\right) ,\text{ \ }i\geq0,
\end{equation*}%
\begin{equation*}
G_{i}=\left( -U_{i+1}^{-1}\right) T_{i+1,i},\text{ \ }i\geq1.
\end{equation*}

By using the $R$-, $U$- and $G$- measures, the UL-type RG-factorization of
the QBD process $T$ is given by
\begin{equation*}
T=\left( I-R_{U}\right) U_{D}\left( I-G_{L}\right) ,
\end{equation*}
where%
\begin{equation*}
U_{D}=\text{diag}\left( {{U_{0}},{U_{1}},{U_{2}},\ldots}\right) ,
\end{equation*}%
\begin{equation*}
R_{U}=\left( {%
\begin{array}{ccccc}
0 & R_{0} &  &  &  \\
& 0 & R_{1} &  &  \\
&  & 0 & R_{2} &  \\
&  &  & \ddots & \ddots%
\end{array}
}\right) ,
\end{equation*}%
\begin{equation*}
\text{ }G_{L}=\left( {%
\begin{array}{ccccc}
0 &  &  &  &  \\
G_{1} & 0 &  &  &  \\
& G_{2} & 0 &  &  \\
&  & G_{3} & 0 &  \\
&  &  & \ddots & \ddots%
\end{array}
}\right) {.}
\end{equation*}

We write
\begin{equation*}
{X}_{k}^{\left( l\right) }=R_{l}R_{l+1}\cdots R_{l+k-1},\text{ \ }%
k\geq1,l\geq0,
\end{equation*}%
\begin{equation*}
Y_{k}^{\left( l\right) }=G_{l}G_{l-1}\cdots G_{l-k+1},\text{ \ }1\leq k\leq
l,
\end{equation*}
and
\begin{equation*}
U_{D}^{-1}=\text{diag}\left( U_{0}^{-1},U_{1}^{-1},U_{2}^{-1},\ldots\right) .
\end{equation*}

By using the UL-type RG-factorization, we obtain%
\begin{equation*}
T_{\max}^{-1}=\left( I-G_{L}\right) ^{-1}U_{D}^{-1}\left( I-R_{U}\right)
^{-1},
\end{equation*}
where%
\begin{equation*}
\left( I-R_{U}\right) ^{-1}=\left( {%
\begin{array}{ccccc}
I & {X_{1}^{\left( 0\right) }} & {X_{2}^{\left( 0\right) }} & {X_{3}^{\left(
0\right) }} & \cdots \\
& I & {X_{1}^{\left( 1\right) }} & {X_{2}^{\left( 1\right) }} & \cdots \\
&  & I & {X_{1}^{\left( 2\right) }} & \cdots \\
&  &  & I & \cdots \\
&  &  &  & \ddots%
\end{array}
}\right) ,
\end{equation*}%
\begin{equation*}
\left( I-G_{L}\right) ^{-1}=\left( {%
\begin{array}{ccccc}
I &  &  &  &  \\
Y_{1}^{\left( 1\right) } & I &  &  &  \\
Y{_{2}^{\left( 2\right) }} & {Y_{1}^{\left( 2\right) }} & I &  &  \\
{Y_{3}^{\left( 3\right) }} & {Y_{2}^{\left( 3\right) }} & {Y_{1}^{\left(
3\right) }} & I &  \\
\vdots & \vdots & \vdots & \vdots & \ddots%
\end{array}
}\right) .
\end{equation*}

Let
\begin{equation*}
T_{\max}^{-1}=\left(
\begin{array}{cccc}
V_{0,0} & V_{0,1} & V_{0,2} & \cdots \\
V_{1,0} & V_{1,1} & V_{1,2} & \cdots \\
V_{2,0} & V_{2,1} & V_{2,2} & \cdots \\
\vdots & \vdots & \vdots & \ddots%
\end{array}
\right) .
\end{equation*}
Then by using $T_{\max}^{-1}=(I-G_{L})^{-1}U_{D}^{-1}(I-R_{U})^{-1}$, we
obtain

\begin{equation*}
V_{m,n}=\left\{
\begin{array}{l}
U_{m}^{-1}Y_{m-n}^{\left( m\right) }+\sum\limits_{i=1}^{\infty}X_{i}^{\left(
m\right) }U_{i+m}^{-1}Y_{i+m-n}^{\left( i+m\right) }{,}  \quad\quad  0\leq n\leq m-1,
\\
U{_{m}^{-1}+\sum\limits_{i=1}^{\infty}}X{{_{i}^{\left( m\right) }}}U{{%
_{i+m}^{-1}}}Y{{_{i}^{\left( i{+m}\right) },}} \quad\quad n=m, \\
X{_{n-m}^{\left( m\right) }}U{_{n}^{-1}+\sum\limits_{i=n-m+1}^{\infty}}X{{%
_{i}^{\left( m\right) }}}U{{_{i+m}^{-1}}}Y{{_{i-n+m}^{\left( {i+m}\right) },}%
}  \quad\quad n\geq m+1.%
\end{array}
\right.
\end{equation*}

The average confirmation time of the arriving internal tip A in the DAG-based
blockchain system is given by
\begin{align*}
E\left[ W_{A}\right] & =-\theta T_{\max}^{-1}e=-\theta\left( I-G_{L}\right)
^{-1}U_{D}^{-1}\left( I-R_{U}\right) ^{-1}e \\
& =-\sum\limits_{i=1}^{\infty}\sum\limits_{j=0}^{\infty}\theta_{i}V_{i-1,j}e.
\end{align*}

\begin{Rem}
In the blockchain system, the average confirmation time of a new transaction is a key performance measure, but its analysis is always very difficult and challenging, e.g., see Li et al. \cite{Li:2018, Li:2019}. For the complicated DAG-based blockchain system, this paper finds a new effective method to be able to numerically compute the average confirmation time of an arriving internal tip by means of the RG-factorization.
\end{Rem}

\section{Numerical Examples}\label{sec:numerical}
In this section, we use numerical examples to check the validity of our
theoretical results, and indicate how some key system parameters influence
the performance measures of the DAG-based blockchain system. To do this, our
analysis is to focus on two issues: (1) Steady-state performance indicators
of the DAG-based blockchain system. (2) The average confirmation time $E\left[
W_{A}\right] $ of the arriving internal tip A.

\subsection{Steady-state performance measures}

Firstly, we discuss how the steady-state performance measures of the DAG-based
blockchain system are affected by the three key system parameters: $\alpha$,
$\mu$ and $\lambda$.

\begin{figure}[ptbh]
\centering \includegraphics[width=10cm]{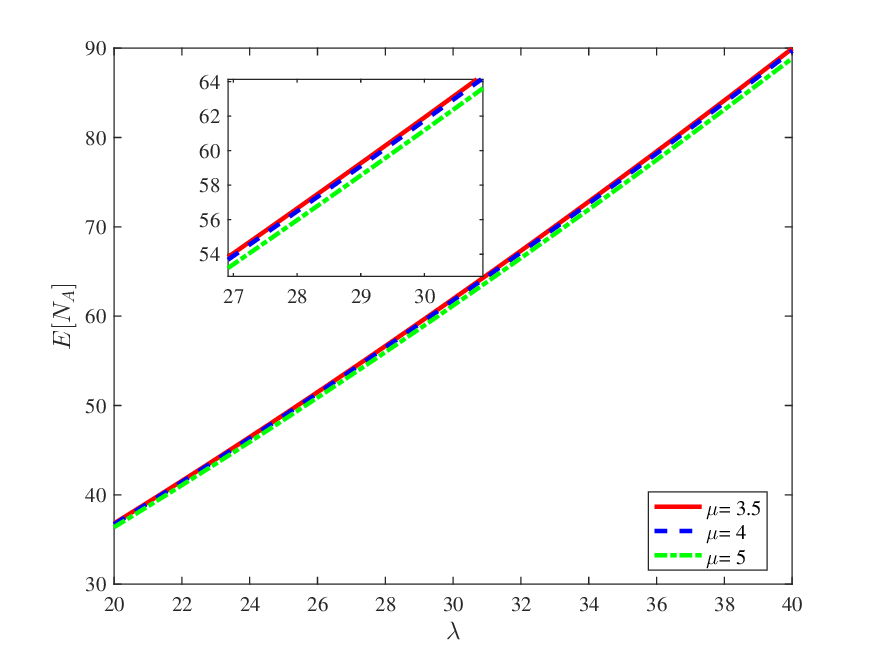}
\caption{$E[N_{A}]$ vs $\lambda$ and $\mu$}
\label{figure:fig-4}
\end{figure}

\begin{figure}[ptbh]
\centering \includegraphics[width=10cm]{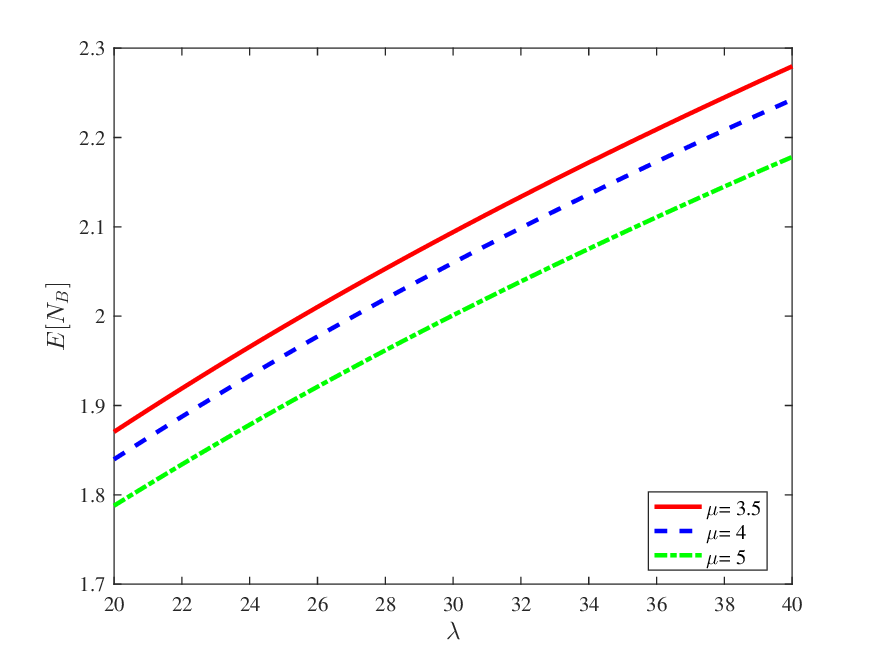}
\caption{$E[N_{B}]$ vs $\lambda$ and $\mu$}
\label{figure:fig-5}
\end{figure}

In Fig. \ref{figure:fig-4} and Fig. \ref{figure:fig-5}, we take that $M=100$, $\alpha=0.45$, $\lambda\in
\lbrack20,40]$, and $\mu=3.5,$ $4,$ $5$. From Fig. \ref{figure:fig-4}, it is easy to see that $E[N_{A}]$ increases as $\lambda$
increases, while $E[N_{A}]$ decreases as $\mu$ increases. From Fig. \ref{figure:fig-5}, we observe how $E[N_{B}]$ changes as $\lambda$ or $\mu$ increases. It can be
seen that $E[N_{B}]$ increases as $\lambda$ increases, while $E[N_{B}]$ decreases as $\mu$ increases.

\textbf{A coupling analysis}: Note that the two numerical results can be intuitively understood by means of a coupling method. As $\lambda$ increases, more and more internal tips (or new transactions) quickly enter the DAG-based blockchain system. Thus, this makes $E[N_{A}]$ and $E[N_{B}]$ increase. On the other hand, when $\alpha$ is constant and as $\mu$ increases, more and more internal tips quickly leave the set of internal tips. Further, note that the tip number in the set of boundary tips decreases due to the connection of each internal tip with two boundary tips, this leads to the decrease of $E[N_{A}]$ and $E[N_{B}]$ accordingly.

\begin{figure}[ptbh]
\centering \includegraphics[width=10cm]{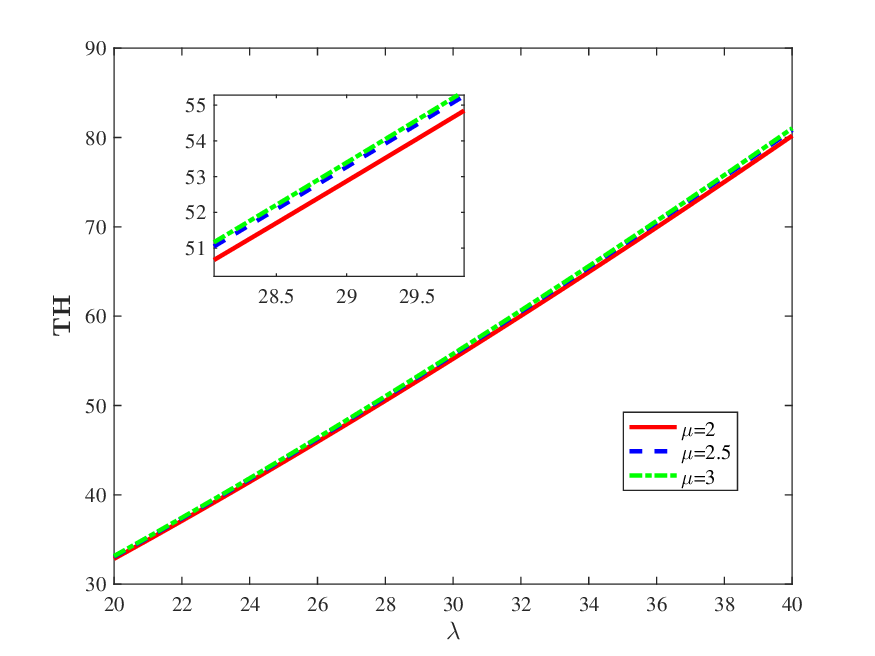}
\caption{TH vs $\lambda$ and $\mu$}
\label{figure:fig-6}
\end{figure}

In Fig. \ref{figure:fig-6}, we explore how \text{TH} is influenced by
parameters $\lambda$ and $\mu $. To this end, we take that $M=100$, $\alpha=0.45$,
$\lambda\in\lbrack20,40]$, and $\mu=2,$ $2.5,$ $3$. From Fig. \ref{figure:fig-6},
it is observed that \text{TH} increases as $\lambda$ or $\mu$ increases. This indicates
that as $\lambda$ and $\mu$ increase, more and more network nodes (or blocks) quickly enter the DAG-based blockchain system, this leads to the increase of \text{TH}.
For this result, we can give an intuitive explanation, as the results are in Fig. \ref{figure:fig-4} and Fig. \ref{figure:fig-5}.
From Fig. \ref{figure:fig-4} and Fig. \ref{figure:fig-5}, we can know that as $\lambda$ increases,
more and more internal tips (or new transactions) can become boundary tips,
so more and more boundary tips have the opportunity to become network nodes
 (or blocks), this makes \text{TH} increases. In addition, as $\mu$ increases, more and more boundary tips become network nodes (or blocks),
 hence this leads to increase of \text{TH}. The numerical results are consistent with our intuitive understanding.

\subsection{The average confirmation time of the arriving internal tip A}

In this subsection, we analyze how the average confirmation time of the arriving internal tip A is
affected by the three key system parameters: $\alpha$, $\mu$, and $\lambda$. In this part, we let $M=50$ for all the numerical examples.

\begin{figure}[ptb]
\centering \includegraphics[width=10cm]{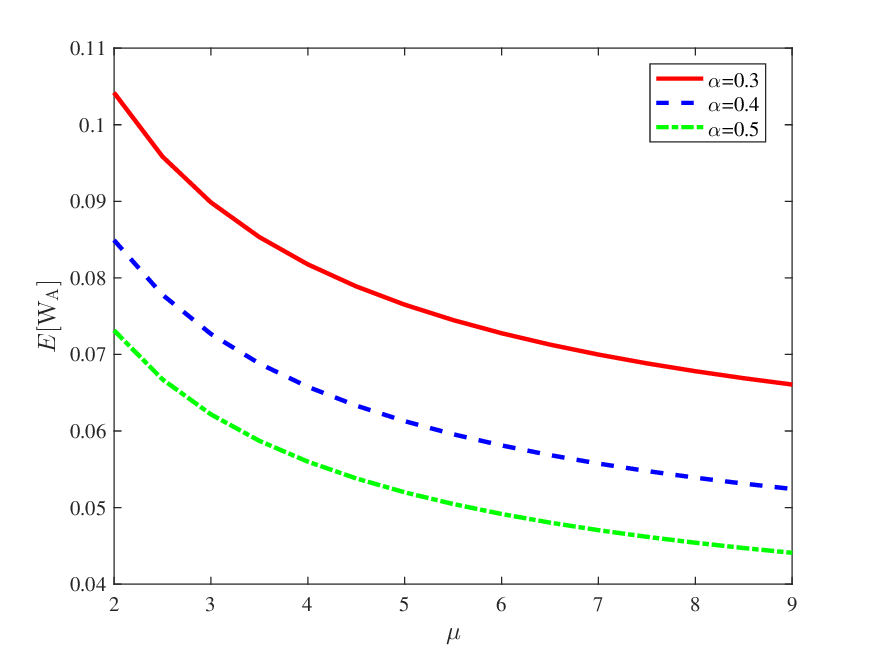}
\caption{$E[W_{A}]$ vs $\mu$ and $\alpha$}
\label{figure:fig-7}
\end{figure}

In Fig. \ref{figure:fig-7}, we observe how $E[W_{A}]$ is influenced by parameters: $\alpha$ and $\mu$. To this end, we take that $\alpha=0.3,0.4,0.5$, $\lambda=30$, and
$\mu\in\left[ 2,9\right] $. It is easy to see from Fig. \ref{figure:fig-7} that $E[W_{A}]$ decreases as $\mu$ or $\alpha$ increases. This indicates that as $\mu$ or $\alpha$ increases, that is, the rate that tips become network nodes (or blocks) increases, then the probability of the arriving internal tip A being approved or connected increases, this makes the average confirmation time of the arriving internal tip A decreases, which is consistent with our intuitive understanding. Furthermore, this result also indicates that the impatient behavior introduced by us can prevent some tips from staying for too much time in the system.

\begin{figure}[ptbh]
\centering \includegraphics[width=10cm]{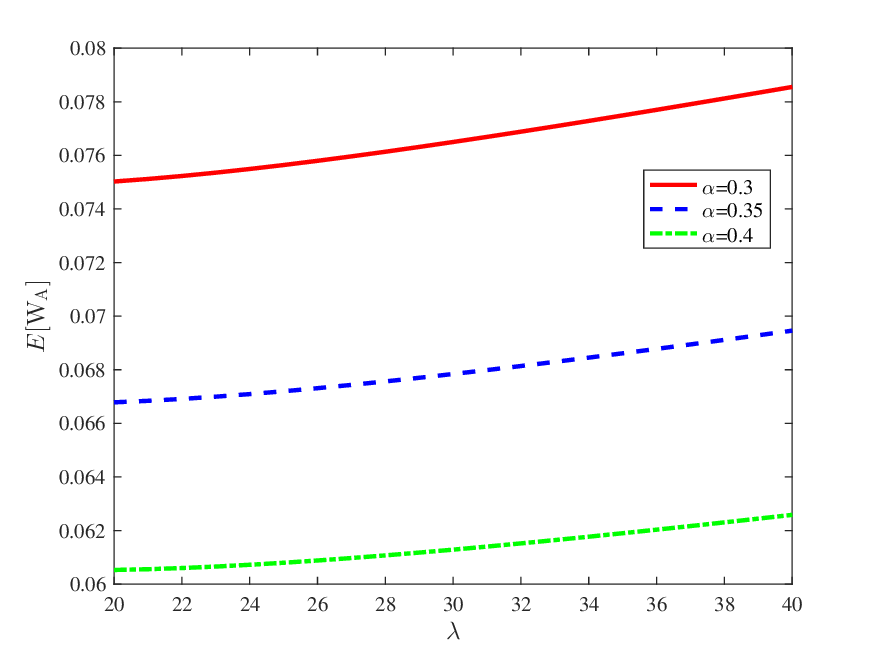}
\caption{$E[W_{A}]$ vs $\lambda$ and $\alpha$}
\label{figure:fig-8}
\end{figure}

In Fig. \ref{figure:fig-8}, we explore how $E[W_{A}]$ is influenced by parameters $\alpha$ and $\lambda$. To this end, we take that $\mu=5$, $\alpha=0.3,0.35,0.4$, $\lambda\in\left[20,40\right] $. It is easy to see from Fig. \ref{figure:fig-8}
that $E[W_{A}]$ increases as $\lambda$ increases, while it decreases as $\alpha$ increases.
The results indicate that as $\lambda$ increases, $E[N_{A}]$ and $E[N_{B}]$ increase based
on the results in Fig. \ref{figure:fig-4} and Fig. \ref{figure:fig-5}.
This causes to decrease the probability of the arriving internal tip A being
approved or connected, and then the average confirmation time of the arriving internal
tip A increases. On the other hand, as $\alpha$ increases, the rate that tips become network
nodes (or blocks) increases, then the probability of the arriving internal tip
A being approved or connected increases. This leads to the average confirmation time of the arriving
internal tip A decreases. The numerical results are also consistent with our intuitive sense.

\begin{figure}[ptbh]
\centering \includegraphics[width=10cm]{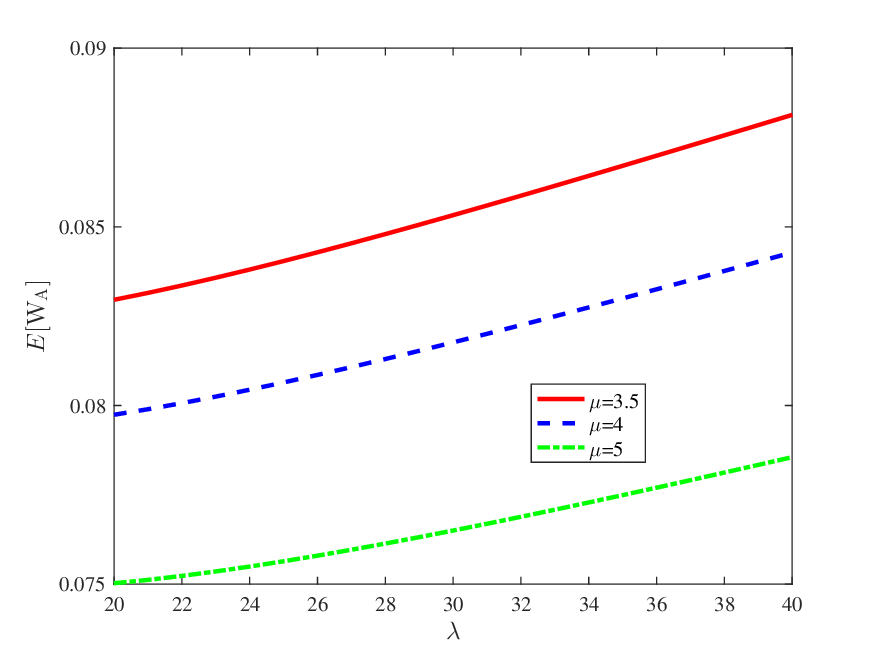}
\caption{$E[W_{A}]$ vs $\lambda$ and $\mu$}
\label{figure:fig-9}
\end{figure}

In Fig. \ref{figure:fig-9}, we indicate how $E[W_{A}]$ is affected by the two
parameters $\mu$ and $\lambda$. To this end, we take that $\alpha=0.3$, $\mu=3.5,4,5$ and $\lambda
\in\lbrack20,40]$. From Fig. \ref{figure:fig-9}, we can see that $E[W_{A}]$
increases as $\lambda$ increases, this is consistent with the results in Fig. \ref{figure:fig-8}.
At the same time, we can see from Fig. \ref{figure:fig-9} that as $\mu$ increases,
$E[W_{A}]$ decreases as $\mu$ increases. This result indicates that as $\mu$ increases,
the rate at which internal tips become boundary tips and network nodes (or blocks) increases,
then the probability of the arriving internal tip A being approved or connected increases.
 These lead to the decrease of the average confirmation time of the arriving internal tip A.
 This numerical result is also consistent with our intuition.

\section{Concluding Remarks}\label{sec:remarks}

Since Bitcoin was proposed by Nakamoto \cite{Nak:2008} in 2008, the serial
structure of blockchain has achieved rapid development. Important examples
include Bitcoin, Litecoin, Ethereum and so on, and the consensus mechanisms
such as PoW, PoS, DPoS and PBFT are widely used. However, such a serial
structure has a
number of essential pitfalls, for example, poor performance and scalability,
limited transaction throughput, high transaction cost, long confirmation
delay, huge energy expenditure, and so forth.

To overcome these pitfalls and drawbacks of blockchain, a data network
structure was found in the DAG-based blockchain with IOTA Tangle. From such
a network perspective, it is well-known that the analysis of the DAG-based
blockchain systems becomes interesting but difficult and challenging. To
investigate the DAG-based blockchain, so far, the simulation models have been
adopted widely while the mathematical modeling and analysis is still scarce in the recent literature. To our best
knowledge, this paper may be the first to set up theory of Markov processes in the study of DAG-based blockchain
systems.

In this paper, we first describe a simple Markov model for the DAG-based
blockchain with IOTA Tangle. Then we set up a continuous-time
level-dependent QBD process to analyze the DAG-based blockchain system. We show
that the QBD process must be irreducible and positive recurrent. By using
the stationary probability vector of the QBD process, we provide a
performance analysis of the DAG-based blockchain system. Furthermore, we propose a new
effective method for computing the average confirmation time of any arriving internal tip
at this system. Crucially, for the complicated DAG-based blockchain system, this paper provides a new effective computational method to deal with the two key performance measures: Throughput and the confirmation time of an arriving internal tip. Finally, we use numerical examples to check the validity of
our theoretical results and indicate how some key system parameters
influence the performance measures of this system.

Note that the Markov process theory opens up a new venue to the study of DAG-based
blockchain systems. Thus, we believe that the methodology and results developed
in this paper can be applicable to deal with more general DAG-based blockchain systems
and open a series of potentially promising research. Along these lines, we
will continue our future research in the following directions:

-- Considering the DAG-based blockchain systems with IOTA Tangle. From $2$
connections to $m$ connections of the boundary tips, we observe how the
multiple connections will affect the performance of the DAG-based blockchain system.

-- Setting up a block-structure Markov process for the DAG-based blockchain
systems and develop an effective algorithm for
computing the matrix-analytic solution. Using the stationary probability
vector, we can analyze the performance of the DAG-based blockchain systems.

-- Developing the fluid and diffusion approximation for analyzing the DAG-based
blockchain systems, when there are several general random factors.

-- Further developing stochastic optimization and dynamic control of the DAG-based
blockchain systems, for example, Markov decision processes, stochastic game,
evolutionary game and so forth.

\section{Acknowledgements}

Quan-Lin Li was supported by the National Natural Science Foundation of
China under grants No. 71671158 and 71932002.

\end{document}